\documentclass[3p,review,12pt]{elsarticle}

\usepackage{amssymb}
\usepackage{lineno}
\usepackage[driverfallback=dvipdfm,colorlinks,bookmarksopen,bookmarksnumbered,citecolor=blue,urlcolor=red]{hyperref}
\usepackage{float} % pour en incluant une figure
\usepackage{color}%\textcolor[rG-B]{0.00,0.00,1.00}{}
\usepackage{epstopdf}
\usepackage{amsfonts}
\usepackage{amsmath}
\usepackage{pdflscape}
\usepackage{graphicx}
\usepackage{subcaption}
\usepackage{soul}
\usepackage{xcolor}
\usepackage{bbm,bm}
\usepackage{stmaryrd}
\usepackage{color}
\usepackage{booktabs}
\usepackage{yhmath}
\usepackage{ulem} %% using delete symbol
\usepackage{flexisym}
\usepackage{adjustbox}
\usepackage{physics}
\usepackage{boldline,multirow}
\graphicspath{{./figures/}}
\usepackage{tabularx}
\usepackage{upgreek}
\usepackage[linesnumbered,ruled,vlined]{algorithm2e}
\usepackage[labelsep=period]{caption}

\usepackage{etoolbox}

\makeatletter
\patchcmd{\MaketitleBox}
  {\vskip36pt}
  {\vskip12pt}
  {}{}
\makeatother
\usepackage{placeins}
\usepackage{xcolor}

\journal{xxx}
\begin{document}
\begin{frontmatter}

\title{Physics-informed latent conditional flow matching for generative modeling of granular flows on inclined planes}

\author[1]{Ziqi Han}
\author[1]{Kuang Shi}
\author[2]{Xuyang Li}
\author[3]{Teng Man}
\author[1]{Yimin Lu\corref{cor}}
\ead{yimin.lu@ttu.edu}

\cortext[cor]{Corresponding author}
\address[1]{Department of Civil, Environmental, and Construction Engineering, Texas Tech University, Lubbock, TX 79409, USA}
\address[2]{School of Construction, University of North Carolina at Charlotte, Charlotte, NC 28223, USA}
\address[3]{College of Civil Engineering, Zhejiang University of Technology, Hangzhou, Zhejiang 310023, China}

%=============================================================================
\begin{abstract}
The internal kinematic and mechanical states of granular flows are difficult to observe or measure, limiting the interpretation of internal particle interactions and their role in flow evolution. In this study, we propose a physics-informed latent conditional flow matching (PI-LCFM) framework for reconstructing complete multivariable flow sequences from partial boundary observations. A variational autoencoder and temporal flow-matching model learn a reusable latent prior over velocity, granular temperature, stress, and solid volume fraction. During inference, boundary-velocity or basal-stress observations and mass- and momentum-conservation constraints guide the latent trajectory through the decoded physical fields, without retraining either model. The results demonstrate that the model accurately reconstructs full spatiotemporal fields from observations containing only 0.3\% of the complete field information, extrapolates beyond the training ranges of inclination and particle number, and recovers key trends in runout scaling, energy evolution, and stress profiles. Compared with a frame-by-frame conditional flow-matching baseline, PI-LCFM reduces the mean velocity-component root-mean-square error by 45.2\%, while observation-guided reconstruction of 500-frame sequences achieves an approximately 44-fold speedup. Physics guidance further improves conservation consistency, supporting the broader potential of this framework for inferring hidden states in particulate systems, with potential applications envisioned in geophysical hazard assessment and industrial process monitoring.
\end{abstract}


\end{frontmatter}

%=============================================================================
\section{Introduction}

Granular flows are widely encountered in industrial processes and geophysical hazards, including particle transport, feedstock handling, landslides, rock avalanches, and debris flows \cite{tahmasebi2023state,lu2026hydro,wang2025numerical,jin2022fidelity}. Understanding their internal kinematics and mechanical responses is essential for interpreting flow evolution and evaluating associated hazards.
However, the internal states of dense granular flows are difficult to access directly in experiments. For example, in inclined plane or flume experiments, measurements are generally limited to the free surface, sidewalls, basal boundaries, or a small number of sensor locations \cite{li2025evolution,trewhela2021conveyor,carleo2019volume,li2022insight}.
Consequently, internal velocity fluctuations, stress transmission, and other mechanical quantities remain largely unobservable \cite{wu2025unified,zhou2025rate}. This limited visibility obscures the mechanisms and compromises the assessment of runout and impact loads in geophysical hazards and predictions of disruptive blockages in industrial handling systems \cite{sun2026deciphering}. Inferring these inaccessible quantities from limited measurements is thus an important inverse problem for understanding granular-flow mechanics and supporting reliable hazard assessment and process design.

Numerical simulations and conventional data-driven models provide complementary approaches to resolving the internal dynamics of granular materials. Particle-scale approaches such as the discrete element method (DEM) explicitly resolve particle motion and contact interactions \cite{shi2022recent,deak2025high,sun2027length}, while coupling DEM with computational fluid dynamics enables the modeling of particle--fluid interactions and hydro-mechanical failures \cite{shan2014coupled,zhang2024microscopic,chen7286222hybrid}. Continuum-based methods, including finite-element \cite{zheng2015modelling,lu2021flow}, smoothed particle hydrodynamics \cite{nguyen2017new}, and the material point method \cite{haeri2022three}, describe bulk motion through governing equations and constitutive relations \cite{kamrin2024advances,xue2023nonlocal}. 
Although these approaches provide detailed mechanical information, repeated parameter calibration and high-fidelity simulations are computationally expensive for inverse reconstruction.
Reduced-order methods, such as proper orthogonal decomposition \cite{berkooz1993proper} and dynamic mode decomposition \cite{schmid2010dynamic}, can reduce the computational cost by constructing lower-dimensional representations of complex flow fields, but they usually struggle to represent strongly nonlinear and transient processes outside the dominant modes captured by the training data.
Deep learning algorithms provide nonlinear alternatives for field reconstruction \cite{fukami2021global,clark2022reconstructing} and surrogate modeling \cite{dubois2022machine,el2025machine}. However, conventional deterministic models produce a single prediction and do not directly represent the non-uniqueness associated with sparse, noisy, or incomplete measurements \cite{guo2025advances}. Moreover, models trained for fixed observation inputs may also require redesign or retraining when the measured quantities or sensor configurations change, while maintaining physical consistency remains challenging.

Generative models provide a probabilistic alternative to deterministic models by representing multiple physical states compatible with incomplete measurements. For fluid-field reconstruction, Shu et al.~\cite{shu2023physics} used a physics-informed diffusion model to recover turbulent fields from sparse measurements or low-resolution inputs, while Gao et al.~\cite{gao2024bayesian} developed a Bayesian conditional diffusion model for spatiotemporal turbulence generation under different prescribed flow conditions. Du et al. \cite{du2024conditional} further combined neural-field compression with latent diffusion to reconstruct full spatiotemporal flows from sparse sensor data without retraining for each observation configuration.
More recently, flow matching (FM) \cite{lipman2022flow} has emerged as an efficient continuous generative formulation that transports samples from a reference distribution toward the target data distribution through a learned ordinary differential equation. 
Parikh et al. \cite{parikh2026conditional} applied FM to reconstruct near-wall turbulent velocity fluctuations from sparse wall measurements while quantifying predictive uncertainty. 
However, a corresponding framework for jointly reconstructing complete multivariable granular-flow sequences remains underdeveloped. Turbulent-velocity reconstruction formulations cannot be transferred directly to this problem without accounting for frictional contact stresses, evolving packing density, and dissipative particle fluctuations that couple the kinematic and mechanical fields during flow and arrest \cite{kamrin2024advances}. 
Our previous work \cite{li2026generative} established a conditional FM framework for reconstructing internal granular-flow states from side-boundary velocity observations. 
However, frame-by-frame generation and a separately trained observation operator limited its efficiency and flexibility, while joint multivariable sequence generation with physical-conservation guidance remained unaddressed.

To address these limitations, this study develops a physics-informed latent conditional flow matching (PI-LCFM) framework for reconstructing granular-flow sequences from limited boundary observations. 
A variational autoencoder (VAE) \cite{kingma2013auto,rombach2022high} compresses the high-dimensional physical fields, allowing FM to learn their temporal evolution jointly in a compact latent space. 
The pretrained model provides a reusable prior over sequences of velocity, granular temperature, stress, and solid volume fraction. During inference, observation discrepancies and mass- and momentum-conservation constraints are evaluated on the decoded fields, and their gradients are propagated through the fixed decoder to guide the latent trajectory. 
This formulation incorporates physical constraints directly into generation and accommodates either boundary-velocity or basal-stress observations without retraining the VAE or FM model. The framework is evaluated across held-out flow conditions and varying observation coverage and noise levels, using both reconstruction errors and granular-flow physics to assess its predictive fidelity.

The main contributions of this study are as follows:  
\begin{enumerate}
    \item Decoder-mediated guidance extends observation-based reconstruction to include mass- and momentum-conservation constraints during inference, enabling physical consistency without physics-informed retraining.
    \item Joint latent sequence generation replaces frame-by-frame reconstruction \cite{li2026generative}, reducing computational cost while capturing temporal dependencies among coupled granular-flow fields.
    \item A reusable multivariable prior recovers hidden kinematic and mechanical states from either velocity or stress observations, without observation-specific retraining.
    \item The framework demonstrates generalization to held-out flow conditions and accommodates sparse and noisy observations, with fidelity assessed through both field accuracy and granular-flow physics.
\end{enumerate}

%=============================================================================
\section{Problem statement}
This study addresses the spatio-temporal reconstruction of gravity-driven granular flows on inclined planes from limited measurements. The target sequence $X$ comprises velocity, granular temperature, stress components, and solid volume fraction in the spatial field, with various plane inclination angles. The observations can be expressed as $x=\mathcal{H}(X)+\eta$, where $\mathcal{H}$ selects the measured variables and space-time locations, and \(\eta\) represents measurement noise. Available measurements typically cover only selected boundary quantities in limited regions, leaving most interior fields unobserved. These measurements may also be spatially sparse, temporally incomplete, or affected by limited resolution and instrumental noise.

Sparse or noisy observations generally do not uniquely determine the interior flow sequence since different internal states can produce similar measured responses \cite{arridge2019solving,stuart2010inverse}. The objective is to approximate the conditional distribution $p(X | x)$, enabling plausible full-sequence reconstructions without repeated high-fidelity simulations. The unobserved fields are also physically related: velocity, stress, and solid volume fraction evolve under gravitational forcing and frictional interactions, subject to mass and momentum conservation \cite{forterre2008flows}. These relationships provide additional constraints but do not necessarily eliminate the non-uniqueness. The central challenge is to recover plausible interior flow sequences that agree with the available observations while maintaining physical consistency.

%=============================================================================
\section{Physics-informed latent conditional flow matching}
\begin{figure}[h!]
    \centering
    \includegraphics[width=\linewidth]{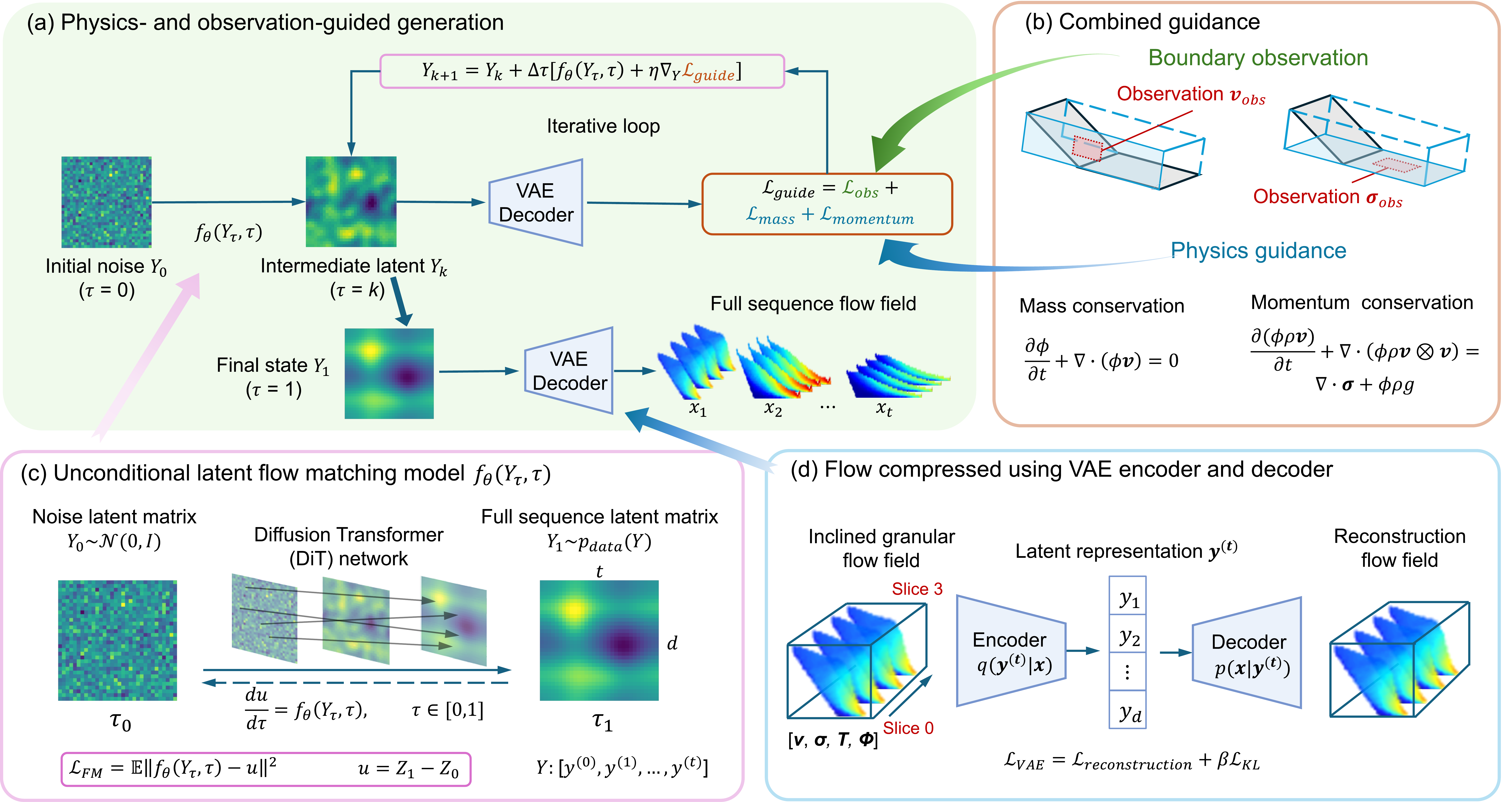}
    \caption{The proposed physics-informed latent conditional flow matching (PI-LCFM) framework. (a) Overall generation process from initial noise to a physically constrained granular flow sequence. (b) Limited boundary observations as the condition for GenAI, with mass and momentum conservations as physics constraints. (c) Unconditional latent flow matching generating a full-sequence latent field from a noise matrix. (d) Latent-space representation of granular-flow sequence using variational autoencoder (VAE). }
    \label{fig:framework}
\end{figure}

Figure \ref{fig:framework} illustrates the proposed PI-LCFM framework for the full-sequence reconstruction of granular-flow fields. The framework combines VAE-based 
dimensionality reduction, unconditional latent flow matching, and training-free guidance from observations and physical constraints. Particle-scale DEM outputs are first coarse-grained onto a fixed Eulerian grid by locally averaging particle kinematics and aggregating particle volumes and contact contributions (more details in Section~\ref{sec:dem_dataset}). The VAE encodes the resulting fields from 4 $x-z$ slices at selected transverse $y$ positions into one compact latent vector at each recorded physical time, referred to as a frame (Figure \ref{fig:framework}(d)). These vectors are assembled into a latent matrix representing the complete temporal sequence. The unconditional FM model learns a continuous vector field that transports Gaussian noise toward the distribution of these latent matrices (Figure \ref{fig:framework}(c)). This formulation enables joint generation of the entire temporal sequence in a reduced-dimensional space.

During inference, an initial Gaussian noise matrix evolves through the learned vector field, with observation and physics guidance applied iteratively (Figure \ref{fig:framework}(a)). At each integration step, the intermediate latent state is decoded into physical fields to evaluate the guidance losses. As illustrated in Figure \ref{fig:framework}(b), observation guidance measures discrepancies from available boundary velocity or stress measurements, while physics guidance penalizes mass- and momentum-conservation residuals. Gradients of the combined loss are propagated through the fixed VAE decoder to correct the latent trajectory. The final latent matrix is decoded to reconstruct the complete flow sequence. Both pretrained modules, VAE and unconditional FM, remain fixed during inference, allowing different observation configurations to be incorporated without retraining.

% -----------------------------------------
\subsection{Latent representation using variational autoencoder network}
The granular-flow state at each physical time step comprises multiple high-dimensional fields. A VAE \cite{kingma2013auto} is employed to compress the multivariable physical fields into a compact latent representation for subsequent sequence modeling. Compared with a deterministic autoencoder, the VAE regularizes the latent representation toward a continuous probabilistic distribution through the Kullback–Leibler (KL) divergence term, producing a smoother and more structured latent space, which is more compatible with the Gaussian reference distribution and continuous probability transport used in flow matching. This structured latent space facilitates subsequent FM-based generative modeling, in which samples are continuously transported from a Gaussian reference distribution toward the latent data distribution. Let \(x_t\) denote the multivariable flow state at time \(t\). The VAE comprises an encoder \(q_{\varphi}(\bm{y}_t \mid \bm{x}_t)\) and a decoder \(p_{\psi}(\bm{x}_t \mid \bm{y}_t)\), parameterized by \(\varphi\) and \(\psi\), respectively.

The encoder represents each input state by a Gaussian latent distribution with mean \(\mu_t\) and standard deviation \(\sigma_t\). A latent vector is sampled using the reparameterization,
\begin{equation}
    y_t=\mu_t+\sigma_t\odot\epsilon,
\qquad \epsilon\sim\mathcal{N}(0,I),
\label{eq:reparameterization}
\end{equation}
where \(\odot\) denotes elementwise multiplication. This formulation allows gradients to propagate through the sampling operation during training. The decoder reconstructs the physical state from the sampled latent vector, $ \hat{x}_t=D_{\psi}(y_t)$.
The VAE is trained by minimizing a reconstruction loss together with a KL divergence term,
\begin{equation}
        \mathcal{L}_{VAE}=\mathcal{L}_{rec}+\beta \mathcal{L}_{KL}
\end{equation}
where \(\mathcal{L}_{rec}\) measures the discrepancy between the input and reconstructed fields. The KL term regularizes \(q_{\varphi}(\bm{y}_t \mid \bm{x}_t)\) toward a standard Gaussian prior, and \(\beta\) controls its contribution to the training objective.

For subsequent FM training, the latent vectors from all \(N_t\) frames of a simulation case are stacked into a sequence matrix:
\begin{equation}
     \boldsymbol{Y}=
 \begin{bmatrix}
    y_1^{\mathrm T}\\
    y_2^{\mathrm T}\\
    \vdots\\
    y_{N_t}^{\mathrm T}
    \end{bmatrix}
    \in\mathbb{R}^{N_t\times d},
\end{equation}
where \(d\) is the latent dimension. Each row represents a single physical time step, while the complete matrix represents the temporal evolution of the flow. The encoder and decoder remain fixed during subsequent FM training and guided inference.

% -----------------------------------------
\subsection{Probabilistic latent flow matching backbone}
The latent FM model generates complete flow sequences in the latent space by learning a continuous transport from a Gaussian reference distribution to the distribution of latent sequence matrices \cite{lipman2022flow}. The learned vector field evolves in generative time $\tau$, which is distinct from the physical time $t$ represented within each sequence. The latent dynamics are governed by:
\begin{equation}
    \frac{d \boldsymbol{Y}_{\tau}}{d\tau}=f_{\theta}(\boldsymbol{Y}_{\tau};\tau),
\end{equation}
where $\boldsymbol{Y}_{\tau}$ denotes the intermediate latent matrix at the generative time $\tau \in[0,1]$, and $f_{\theta}$ is the neural vector field.

Let $\boldsymbol{Y}_1$ denote the latent matrix encoded from a complete granular-flow sequence, and let $\boldsymbol{Y}_0$ denote a random matrix sampled from a standard Gaussian distribution with the same dimensions. Flow matching establishes a continuous probability path between the Gaussian reference distribution at $\tau = 0$ and the latent data distribution at $\tau = 1$. For a linear probability path, the intermediate latent state is defined as
\begin{equation}
   \boldsymbol{Y}_{\tau}=(1-\tau)\boldsymbol{Y}_0+\tau \boldsymbol{Y}_1.
\end{equation}
The corresponding target transport velocity is $\boldsymbol{u}_{\tau}=\boldsymbol{Y}_1-\boldsymbol{Y}_0$. A neural vector field $f_{\theta}(\boldsymbol{Y}_{\tau};\tau)$ is trained to approximate this target velocity by minimizing the difference between the predicted and analytical transport velocities. 
\begin{equation}
\mathcal{L}_{\mathrm{FM}}(\theta)
=
\mathbb{E}_{\substack{
\tau\sim\mathcal{U}(0,1),\\
\boldsymbol{Y}_0\sim p_0,\,
\boldsymbol{Y}_1\sim q_{\mathrm{latent}}
}}
\left[
\left\|
f_{\theta}(\boldsymbol{Y}_{\tau},\tau)
-
\boldsymbol{u}_{\tau}
(\boldsymbol{Y}_{\tau}\mid\boldsymbol{Y}_1)
\right\|_{F}^{2}
\right],
\end{equation}
where $p_{0}$ denotes the standard Gaussian reference distribution, $q_{latent}$ is the distribution of latent-encoded training sequences. The conditional velocity field $\boldsymbol{u}_{\tau}(\boldsymbol{Y}_{\tau}\mid\boldsymbol{Y}_1)$ specifies the transport direction along this path, while $f_{\theta}(\boldsymbol{Y}_{\tau};\tau)$ is trained to approximate this velocity field. During generation, the learned ODE is integrated from $\tau=0$ to $\tau=1$, progressively transforming an initial Gaussian latent matrix into a complete granular flow latent sequence.

The neural vector field is parameterized by a temporal adaptation of the Diffusion Transformer (DiT) architecture \cite{peebles2023scalable}, referred to here as Temporal DiT. Each row of the latent matrix corresponds to the flow state at one physical time step and is projected into a token embedding. Temporal positional embeddings encode the ordering of these tokens, while the generative time \(\tau\) is embedded separately to condition the transformer blocks. Self-attention models dependencies across the temporal sequence, and a final projection maps the token features back to the latent dimension. The network outputs a matrix with the same dimensions as \(Y_\tau\), representing the predicted latent transport velocity \(f_\theta(\boldsymbol{Y}_\tau,\tau)\). The DiT architecture is thus used for FM velocity prediction rather than diffusion-based denoising.

% -----------------------------------------
\subsection{Physics-informed training-free posterior sampling with observation guidance}
The unconditional FM model provides a prior over complete flow sequences but does not incorporate case-specific observations or explicitly impose physical conservation laws. We introduce training-free guidance to incorporate both during inference, with the pretrained model parameters remaining fixed.
From a Bayesian perspective, observations and physical constraints define an energy-weighted target distribution:
\begin{equation}
    \pi(\boldsymbol{X}\mid x)
\propto
p_{\mathrm{prior}}(\boldsymbol{X})
\exp\left[
-\lambda_{\mathrm{obs}}\mathcal{L}_{\mathrm{obs}}(\boldsymbol{X};x)
-\lambda_{\mathrm{phy}}\mathcal{L}_{\mathrm{phy}}(\boldsymbol{X})
\right]
\end{equation}
where \(X\) denotes the complete physical-field sequence and \(x\) denotes the available observations. The observation loss measures discrepancies between the reconstructed and observed quantities, while the physics loss penalizes mass- and momentum-conservation residuals. At each guidance step, the latent estimate is decoded, and the observation loss is the mean squared difference between the decoded predictions and measurements at the observed space-time locations. Unobserved entries do not contribute to this loss. The weights \(\lambda_{\mathrm{obs}}\) and \(\lambda_{\mathrm{phy}}\) control their relative contributions. When the weighted observation loss is defined as a negative log-likelihood, the observation term admits a measurement-likelihood interpretation.

The conservation constraints use the decoded velocity components \(v_x\) and \(v_z\), stress components \(\sigma_{xx}\), \(\sigma_{xz}\), and \(\sigma_{zz}\), and solid volume fraction \(\phi\). They are evaluated independently at each physical time and transverse slice using a two-dimensional formulation.
The bulk density is defined as
\begin{equation}
    \rho=\rho_s\phi,
\end{equation}
where \(\rho_s\) is the particle material density and \(\phi\) is the solid volume fraction. The mass residual is
\begin{equation}
R_{m} =
\frac{\partial(\rho v_x)}{\partial x}
+
\frac{\partial(\rho v_z)}{\partial z}.
\label{eq:mass_residual}
\end{equation}
The two in-plane momentum residuals are
\begin{align}
R_x ={}&
\frac{\partial(\rho v_x^2)}{\partial x}
+
\frac{\partial(\rho v_xv_z)}{\partial z}
-
\frac{\partial\sigma_{xx}}{\partial x}
-
\frac{\partial\sigma_{xz}}{\partial z}
-\rho g_x, \\
R_z ={}&
\frac{\partial(\rho v_xv_z)}{\partial x}
+
\frac{\partial(\rho v_z^2)}{\partial z}
-
\frac{\partial\sigma_{xz}}{\partial x}
-
\frac{\partial\sigma_{zz}}{\partial z}
-\rho g_z.
\label{eq:momentum_residual}
\end{align}

Advective fluxes are evaluated by coarse-grained field, obtained by mapping the particle-scale DEM data onto a structured Eulerian grid through cell averaging (Section~\ref{sec:dem_dataset}). The grid spacings are \(\Delta x=\Delta z=4~\mathrm{mm}\). The conservation constraints are evaluated independently at each recorded frame over interior cells occupied by granular material.
Temporal derivative terms are omitted in the current implementation because the variation of solid fraction against time is relatively small, and the temporal evolution is included in the pretrained FM model. 
Let \(M_{t,s,i,j}\) denote binary mask for physical time $t$, transverse slice \(s\), and cell \((i,j)\). The mask equals one only for interior cells whose own solid volume fraction and those of their four axial neighbors exceed \(\phi_{\min}\),  while all other entries are zero. The threshold is set to $\phi_{\min}=0.0327$, corresponding to the solid volume fraction of a single particle within a grid cell. This excludes empty regions, cells adjacent to the granular free boundary, and the outermost grid cells where centered differences cannot be evaluated. 
The masked residual losses are defined as
\begin{align}
\ell_{mass} &=
\frac{\sum_{n,s,i,j}MR_m^{\,2}}
     {\sum_{n,s,i,j}M},\\
\ell_{momentum} &=
\frac{1}{2}
\left[
\frac{\sum_{n,s,i,j}MR_x^{\,2}}
     {\sum_{n,s,i,j}M}
+
\frac{\sum_{n,s,i,j}MR_z^{\,2}}
     {\sum_{n,s,i,j}M}
\right].
\label{eq:residuals}
\end{align}
The physics-guidance loss is defined as
\begin{equation}
\mathcal{L}_{\mathrm{phy}}
=
\alpha_m \ell_{mass}
+
\alpha_p \ell_{momentum},
\label{eq:physics_loss}
\end{equation}
where \(\alpha_m\) and \(\alpha_p\) weight the mass- and momentum-conservation residuals, respectively; their values are listed in Table~\ref{tab:hyperparams} (\ref{app:hyperparameter_network}). These constraints couple the velocity, stress, and solid volume fraction fields through the mass and momentum balances. Granular temperature is not explicitly included in the present physics guidance since its evolution requires a granular-energy balance accounting for fluctuation-energy production, transport, and dissipation, which are not explicitly modeled in the current formulation. 

To guide generation toward states consistent with these constraints, an additional correction velocity is introduced into the latent dynamics
\begin{equation}
    \frac{d \boldsymbol{Y}_{\tau}}{d\tau}=f_{\theta}(\boldsymbol{Y}_{\tau};\tau)+\boldsymbol{u}_{\tau}'
\end{equation}
At each generative time \(\tau\), the intermediate latent matrix is decoded into physical fields using the fixed VAE decoder. The combined guidance loss is evaluated as
\begin{equation}
    \mathcal{L}_{g}(\boldsymbol{Y}_{\tau};y)
=
\lambda_{\mathrm{obs}}
\mathcal{L}_{\mathrm{obs}}\!\left(D_{\psi}(\boldsymbol{Y}_{\tau});y\right)
+
\lambda_{\mathrm{phy}}
\mathcal{L}_{\mathrm{phy}}\!\left(D_{\psi}(\boldsymbol{Y}_{\tau})\right).
\end{equation}
Its gradient is propagated through the decoder to obtain the latent correction velocity,
\begin{equation}
    \boldsymbol{u}_{\tau}'=-b\left\|f_{\theta}(\boldsymbol{Y}_{\tau},\tau)\right\|_{F} \frac{\nabla_{\boldsymbol{Y}_{\tau}}\mathcal{L}_{g}}{\left\|\nabla_{\boldsymbol{Y}_{\tau}}\mathcal{L}_{g}\right\|_{F}}
\end{equation}
Here, \(b\) controls the correction magnitude relative to the learned FM velocity. For a nonzero guidance gradient, the correction follows the negative gradient of the combined loss, balancing agreement with observations and the imposed physical constraints.
This procedure provides approximate posterior guidance motivated by the target distribution, rather than a guarantee of exact posterior sampling. The VAE and FM parameters remain fixed throughout inference, allowing different observation configurations and physical constraints to be incorporated without retraining either model.

%=============================================================================
\section{Application to inclined granular flows}
\subsection{DEM simulations and dataset construction}
\label{sec:dem_dataset}
\begin{figure} [h!]
    \centering
    \includegraphics[width=\linewidth]{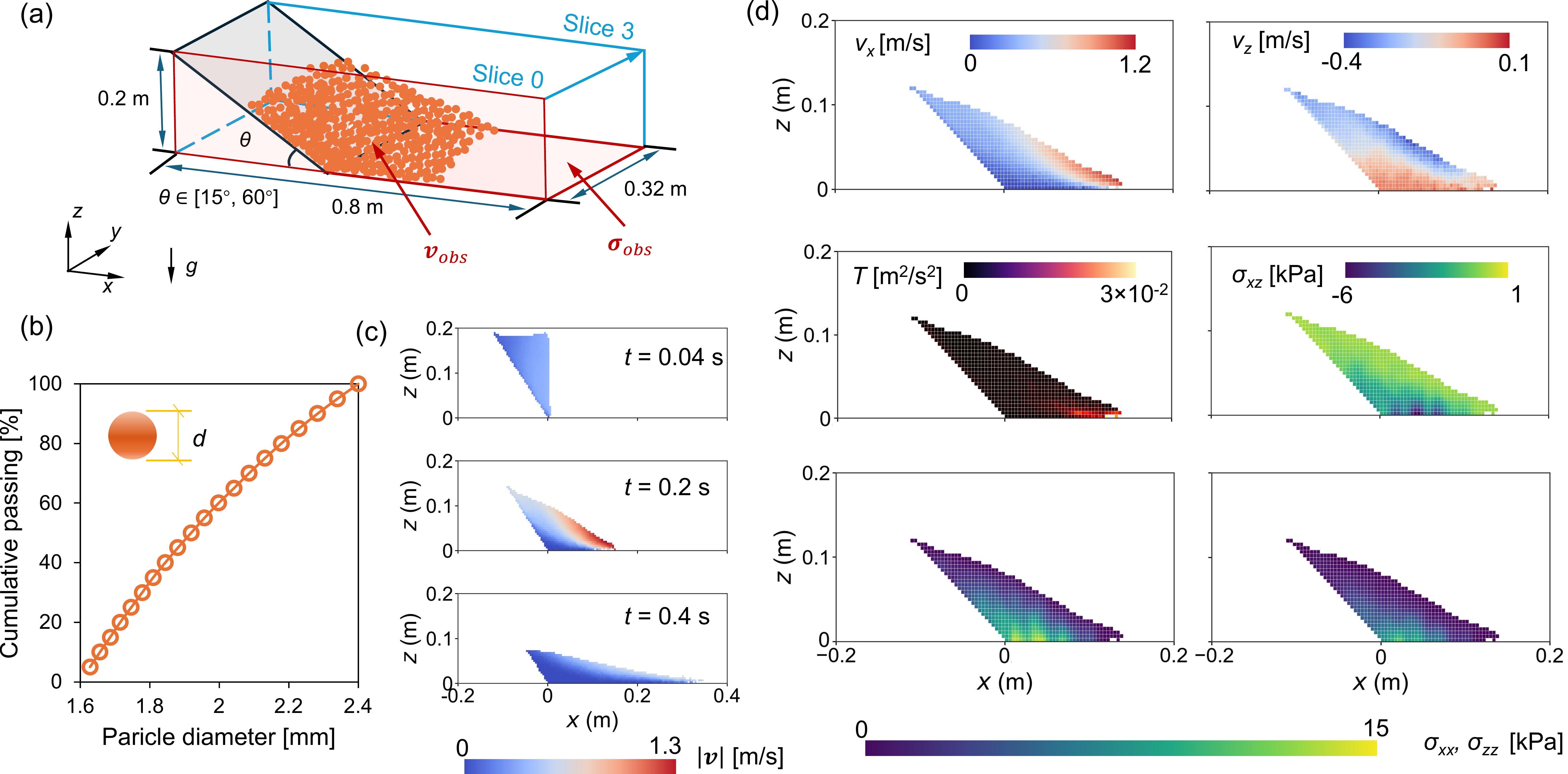}
    \caption{Numerical modeling and construction of the granular-flow dataset. (a) Schematic view of the discrete element method (DEM) computational configuration. (b) Particle size distribution in DEM simulations. (c) Representative temporal evolution from release to deposition. (d) Representative coarse-grained results of velocity, granular temperature, and stress fields.}
    \label{fig:dem_results}
\end{figure}

The dataset was generated through particle-resolved simulations of gravity-driven granular flows using DEM. As illustrated in Figure~\ref{fig:dem_results}(a), the computational domain consists of an inclined plane connected to a horizontal plane. Periodic boundary conditions were imposed in the transverse $y$ direction. The inclination angle was varied from $15^\circ$ to $60^\circ$. During initialization, a retaining plate located at the junction of the two planes ($x=0$) confined the particles on the inclined plane.

A polydisperse particle-size distribution was employed to represent particle-scale variability, as shown in Figure~\ref{fig:dem_results}(b). The particles had a mean diameter of 2 mm, with 20,000$\sim$60,000 spherical particles per simulation. A Hookean linear spring--dashpot contact model was employed \cite{silbert2001granular}, with normal and tangential stiffnesses of $k_n=1\times10^5~\mathrm{N/cm}$ and $k_t=4\times10^4~\mathrm{N/cm}$ respectively, and a restitution coefficient of $e=0.8$. The simulation time step was $1\times10^{-6}$ s, and particle data were recorded every 0.002 s. Further details of the DEM contact formulation are provided in our previous work \cite{li2026generative}.

Before release, the particles were allowed to settle under gravity until a mechanically stable packing was obtained. The retaining plate was then instantaneously removed, allowing the granular material to accelerate along the inclined plane, spread onto the horizontal plane, and eventually deposit. Representative snapshots of this temporal evolution are presented in Figure~\ref{fig:dem_results}(c).
The particle-scale results were coarse-grained onto a structured Eulerian grid to obtain fixed-dimensional continuum fields \cite{nicot2013definition,chung2019rapid,chung2023granular}. The computational domain measures $0.8\times0.32\times0.18$ m in the $x$, $y$, and $z$ directions, respectively, with grid-cell dimensions of $4\times8\times4$ mm. Four \(x\)–\(z\) slices were extracted at selected transverse $y$ positions, each containing $200\times45$ cells. Each frame comprises 7 physical variables on these 4 slices: velocity components \(v_x\) and \(v_z\), granular temperature \(T\), stress components \(\sigma_{xx}\), \(\sigma_{xz}\), and \(\sigma_{zz}\), and solid volume fraction \(\phi\). The frame-level state is written as:
\begin{equation}
{x}_t =
\left[v_x, v_z, T, \sigma_{xx}, \sigma_{xz}, \sigma_{zz}, \phi\right],
\end{equation}
Each entry represents the corresponding field across all 4 slices. The solid volume fraction $\phi$ provides the bulk-density information needed to evaluate the mass- and momentum-conservation constraints.  Representative coarse-grained fields are shown in Figure~\ref{fig:dem_results}(d).

The cell-averaged velocity and granular temperature were calculated from particle velocities following the procedure described in our previous work \cite{li2026generative}. The solid volume fraction was calculated as the total volume of particles whose centers lay within a grid cell divided by the cell volume. Each particle was assigned entirely to the cell containing its center. The cell-scale contact stress was calculated following the microscopic stress formulation \cite{chung2019rapid} as
\begin{equation}
\overline{\sigma}_{ij}
=
\frac{1}{V_{\mathrm{cell}}}
\sum_{k=1}^{N_p}
\sum_{c=1}^{N_c^{(k)}}
r_{c,i}^{(k)}F_{c,j}^{(k)},
\label{eq:cell_contact_stress}
\end{equation}
where $V_{\mathrm{cell}}$ is the cell volume, $N_p$ is the number of particles associated with the cell, $N_c^{(k)}$ is the number of contacts associated with particle $k$, and $r_{c,i}^{(k)}$ and $F_{c,j}^{(k)}$ denote the $i$-component of the branch vector and the $j$-component of the contact force, respectively. To reduce particle-scale fluctuations, the raw stress fields were spatially averaged using a 3 $\times$ 3 cells neighborhood, with boundary treatment following the procedure described in \cite{chung2023granular}. The averaging operation is detailed in \ref{app:coarse_graining}.

A series of simulations was conducted by varying the inclination angle, particle number, and random seed for particle initialization. The final dataset comprised 88 simulation cases, with each case containing 500 flow-field snapshots sampled at intervals of 0.002 s. 

\FloatBarrier

% -----------------------------------------
\subsection{Training and implementation details}

\subsubsection{Data preprocessing and training strategy}
The 88 DEM cases were divided into 73 training cases and 15 test cases, with all frames from each case assigned to the same subset. Selected inclination angles and particle numbers were reserved exclusively for testing to assess generalization to unseen flow conditions, and the details are shown in \ref{app:dataset_conditions}.

The VAE was trained using individual frames as samples, without explicitly modeling their temporal dependence. Within the training cases, 80\% of the frames were used for parameter optimization and the remaining 20\% for validation. After VAE training, the encoder was frozen and used to construct the latent training set for the FM model. To augment the latent training data, 5 independently perturbed versions of each training sequence were generated by adding element-wise Gaussian noise, $\widetilde{Y}=Y+\epsilon$, where $\epsilon_{ij}\sim\mathcal{N}(0,0.05^2)$. The VAE parameters remained fixed throughout FM training.

\subsubsection{Network architecture and hyperparameter configuration}
The VAE uses a convolutional encoder–decoder architecture with a Gaussian latent representation. Each input frame contains 4 spatial slices, each with 7 variables. 4 convolutional layers downsample each slice independently, and two 3D convolutional layers fuse the resulting features across slices. Separate fully connected layers map the flattened features to the latent mean \(\mu_t\) and log variance \(\log\sigma_t^2\) . A latent vector \(y_t\in\mathbb{R}^{d}\ (d=256)\) is then sampled using the reparameterization trick, which is describe in Eq. (\ref{eq:reparameterization}). The decoder reconstructs each slice through 4 upsampling stages.

The latent FM backbone uses a Temporal DiT with 8 transformer blocks. Each row of the latent matrix is linearly projected to a 768-dimensional token and combined with a sinusoidal positional embedding. Each block contains eight-head self-attention and a feed-forward multilayer perceptron (MLP) with an expansion ratio of 4. The generative time \(\tau\) modulates both branches through adaLN-Zero using shift, scale, and gating parameters. A final linear layer projects each token back to 256 dimensions, yielding the predicted latent transport velocity \(f_{\theta}(\boldsymbol{Y}_{\tau},\tau)\).

Both networks were trained using the Adam optimizer \cite{kingma2014adam}, with validation-based early stopping and a maximum of 2,000 epochs. Guided inference used the explicit Euler method to integrate the latent ODE over \(\tau\in[0,1]\). Detailed network configurations and training and inference hyperparameters are provided in Table~\ref{tab:hyperparams} in the \ref{app:hyperparameter_network}.

\subsection{Evaluation metrics}
Reconstruction results were evaluated against the DEM reference fields using the mean absolute error (MAE), root-mean-square error (RMSE), and normalized root-mean-square error (NRMSE). For each variable \(c\), the MAE and RMSE were calculated over \(N\) evaluated spatial and temporal entries:
\begin{equation}
\mathrm{MAE}_{c}=
\frac{1}{N}
\sum_{i=1}^{N}
\left|
\hat{X}_{i}^{(c)}-X_{i}^{(c)}
\right|,
\end{equation}
\begin{equation}
\mathrm{RMSE}_{c}
=
\sqrt{
\frac{1}{N}
\sum_{i=1}^{N}
\left(
\hat{X}_{i}^{(c)}-X_{i}^{(c)}
\right)^2
},
\end{equation}
where \(\hat{X}_{i}^{(c)}\) and \(X_{i}^{(c)}\) denote the predicted and reference values of variable \(c\), respectively. Both metrics retain the physical units of the evaluated variable, with RMSE assigning greater weight to larger errors.

To account for differences in scale among the physical variables, RMSE was normalized by the range of the corresponding DEM reference field \cite{dupuis2018surrogate},
\begin{equation}
\mathrm{NRMSE}_{c}
=
\frac{\mathrm{RMSE}_{c}}{\max\big(X^{(c)}\big)-\min\big(X^{(c)}\big)}
\end{equation}
Lower metric values indicate closer agreement with the DEM reference fields.

%=============================================================================
\section{Results}
This section first examines unconditional sequence generation by the pretrained model, followed by reconstruction from partial observations. Reconstruction accuracy and physical consistency are then evaluated through several metrics.

\begin{figure}[h!]
    \centering
    \includegraphics[width=\linewidth]{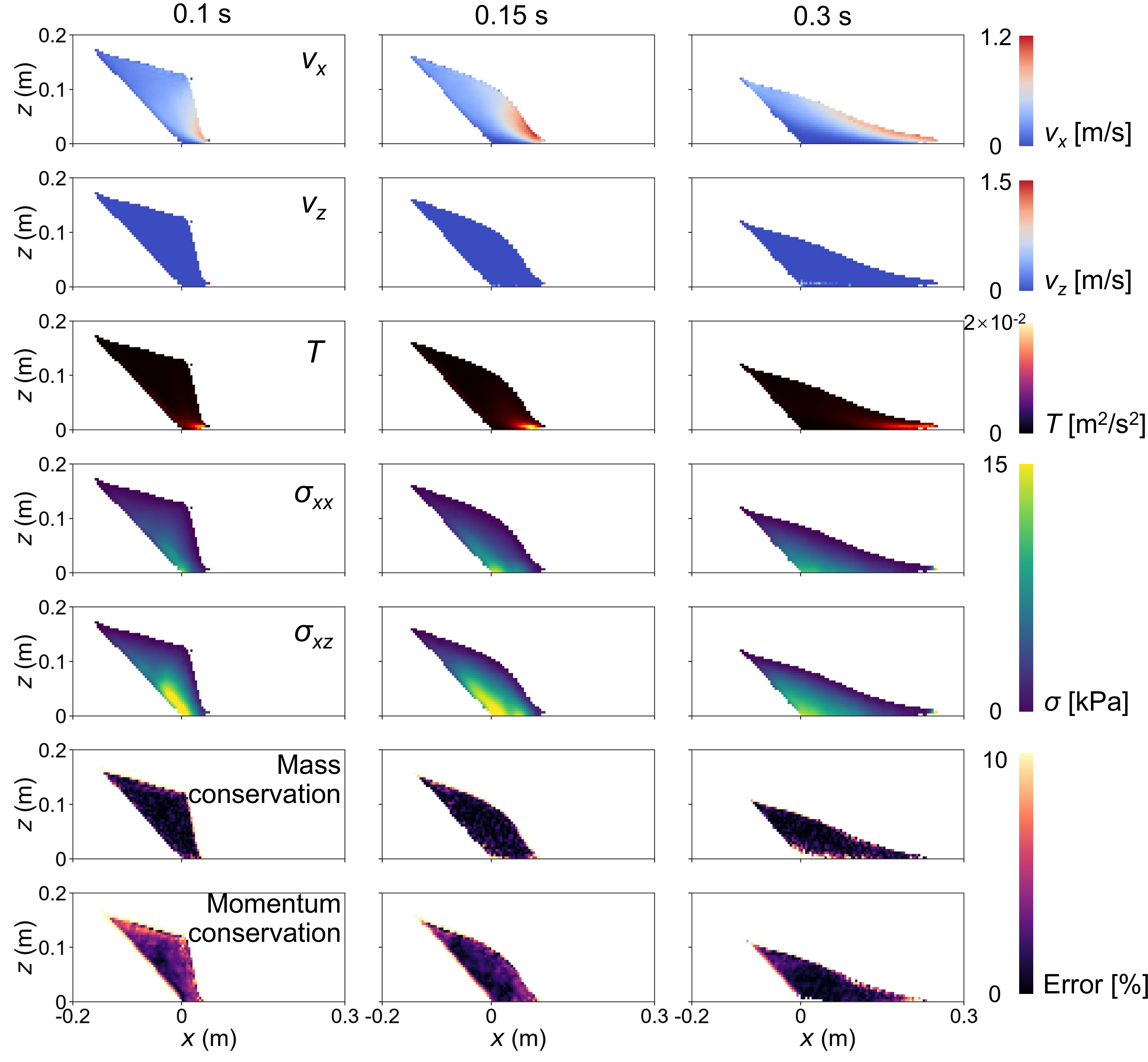}
    \caption{Representative unconditional generation results. Columns correspond to \(t=0.1\), \(0.15\), and \(0.3\) s. Rows show horizontal velocity \(v_x\), vertical velocity \(v_z\), granular temperature \(T\), stress components \(\sigma_{xx}\) and \(\sigma_{xz}\), and mass- and momentum-conservation residuals.}
    \label{fig:uncondition_gen}
\end{figure}

% -----------------------------------------
\subsection{Unconditional spatiotemporal sequence generation}
Complete granular-flow sequences were generated from Gaussian latent noise using the pretrained FM model, without observation or physics guidance. Figure \ref{fig:uncondition_gen} shows representative generated fields at \(t=0.10\), \(0.15\), and \(0.30\) s, together with mass- and momentum-conservation residuals evaluated on the decoded fields.

The generated sequences capture the main spatiotemporal features of the velocity, granular temperature, and stress fields without observation or physics guidance. These results support the use of the pretrained model as a generative prior for conditional reconstruction. Conservation residuals remain relatively small over most of the domain, with the largest deviations concentrated near the flow boundary at early times.
\FloatBarrier

% -----------------------------------------
\subsection{Conditional reconstruction from partial boundary observations}
Conditional reconstruction was evaluated using two observation configurations: partial side-boundary velocity measurements and partial basal stress measurements. The velocity observations comprised \(v_x\) and \(v_z\) within a selected region of slice 0, and the stress observations comprised \(\sigma_{xz}\) and \(\sigma_{zz}\) within a selected basal region. In both configurations, the same pretrained VAE and FM models were used to reconstruct the complete multivariable flow sequence from limited observations. In each configuration, approximately 0.3\% of the complete multivariable field data were used to reconstruct the full flow sequence. The observation fraction is defined as the number of observed scalar entries divided by the total number of scalar entries across all seven variables, four slices, and recorded time steps.

\begin{figure}[h!]
    \centering
    \includegraphics[width=\linewidth]{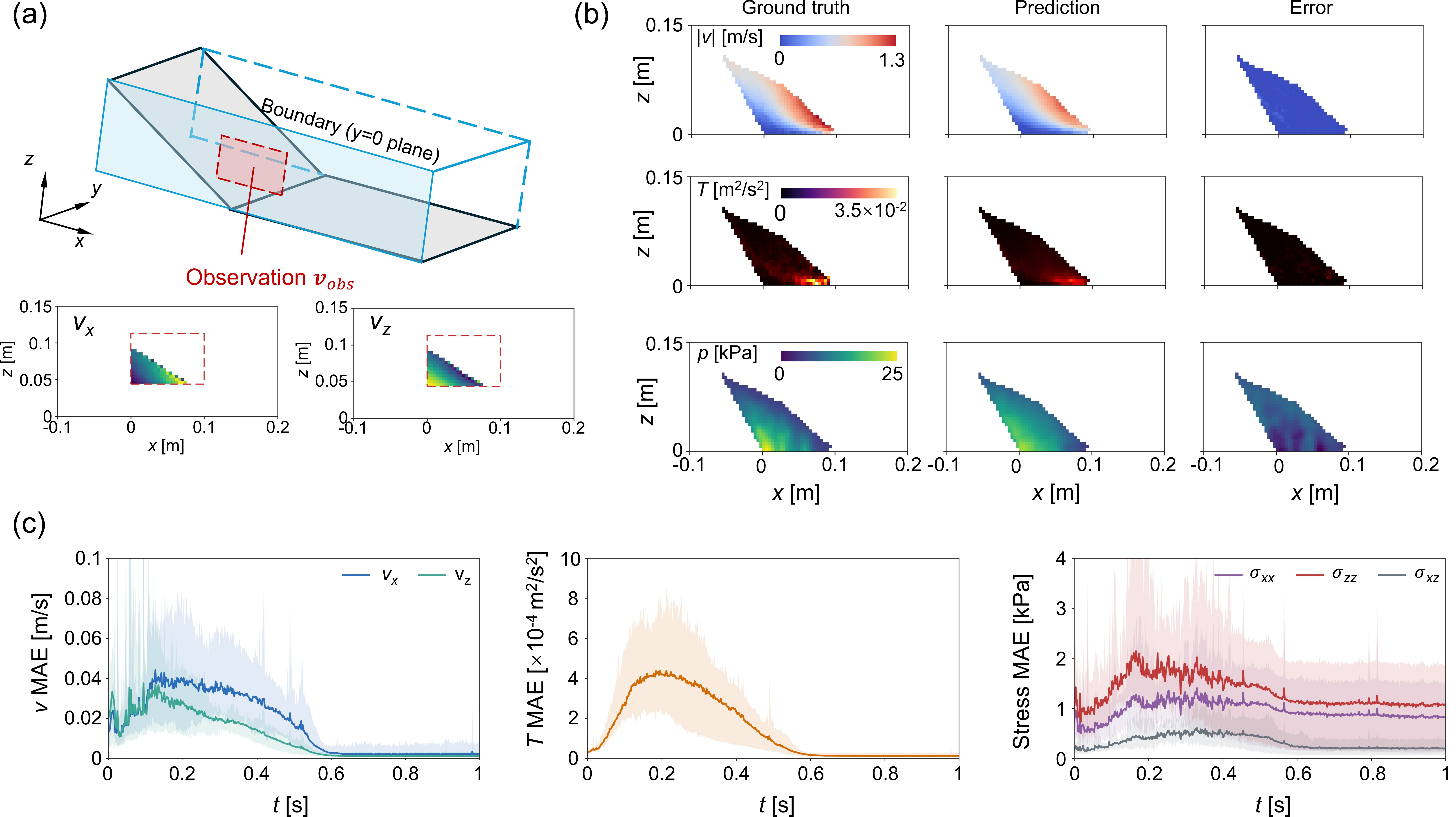}
    \caption{Conditional generation from partial boundary velocity observations. (a) Schematic view of partial velocity observation: two velocity components $v_x$ and $v_z$ are observed within a rectangular window (red dashed box) on the slice 0. (b) Ground truth, GenAI prediction, and absolute error contour of the velocity magnitude $|\boldsymbol{v}|$, granular temperature $T$, and pressure $p$. (c) Temporal evolution of the mean absolute error (MAE) of individual velocity components, granular temperature, and stress components; solid lines denote mean over all test cases and shaded bands indicate the range across all test cases.}
    \label{fig:partial_obs_v}
\end{figure}

Figure \ref{fig:partial_obs_v}(a) shows the side-boundary velocity observations used to reconstruct the \(60^\circ\), 20,000-particle case. Both the inclination angle and particle number lie outside their respective training ranges, making this a joint extrapolation test. Figure \ref{fig:partial_obs_v}(b) compares the DEM reference, GenAI prediction, and error fields for velocity magnitude, granular temperature, and pressure at \(t=0.16\) s, demonstrating that the reconstruction captures the main spatial patterns. The pressure \(p\) and shear stress \(q\) were calculated from the stress tensor as defined in \ref{app:stress_definitions}. Granular temperature and pressure are recovered despite neither quantity being directly observed. The MAEs of the three variables were $0.05$ m/s, $8.651\times10^{-4}$ m²/s², and $3.713$ kPa, respectively, and the corresponding RMSEs were $0.06$ m/s, $2.25\times10^{-3}$ m²/s², and $4.535$ kPa.

Figure \ref{fig:partial_obs_v}(c) summarizes the temporal MAE evolutions of the observed and unobserved variables over 500 frames for 15 test cases.  Solid lines indicate the case-averaged MAE, while shaded bands show the minimum–maximum range across cases. Velocity and granular-temperature errors generally decrease after the early transient stage. Averaged over all cases and 500 frames, the MAEs of \(v_x\) and \(v_z\) were $0.017$ and $0.011$ m/s, respectively. The corresponding averages were $1.447\times10^{-4}$ m²/s² for granular temperature and 0.954, 1.333, and 0.322 kPa for \(\sigma_{xx}\), \(\sigma_{zz}\), and \(\sigma_{xz}\), respectively. These results extend the reconstruction assessment to the complete temporal sequences and quantify errors in both observed and unobserved variables.

Partial basal stress observations provide an alternative experimental measurement configuration motivated by experimental studies of granular-flow dynamics \cite{li2025evolution,li2022insight,cagnoli2012granular}. Figure~\ref{fig:partial_obs_sigma} in \ref{app:basal_stress} presents the reconstruction for a representative \(40^\circ\) case at \(t=0.24~\mathrm{s}\), together with temporal error statistics across the 15 test cases. The reconstructed fields capture the main spatial patterns of velocity and granular temperature, although neither quantity is directly observed. Together with the velocity-conditioned results, it is demonstrated that the same pretrained prior can recover internal kinematic and mechanical fields from different observation modalities without retraining.

\FloatBarrier

% -----------------------------------------
\subsection{Evaluation of reconstructed granular flow physics}
The physical characteristics of the PI-LCFM predictions were evaluated at three complementary levels: macroscopic runout scaling, temporal energy evolution, and local continuum-field profiles. These evaluations examine whether the generated sequences reproduce physical relationships that were not explicitly imposed by the observation guidance.

Previous studies have shown that the normalized runout distances of granular flows with different system sizes and inclination angles can be collapsed onto a master curve after accounting for finite-size effects and the effective mobilized length at small inclination angles \cite{chen2026collapse}. Following this scaling relation, the scaled inclination parameter was defined as
\begin{equation}
    X=
\left(\frac{L_i}{D_p}\right)^{1/\nu}
\tan\theta,
\end{equation}
and the scaled normalized runout distance was expressed as,
\begin{equation}
    Y=
\left(\frac{L_i}{D_p}\right)^{\beta_1/\nu}
\left(\frac{\Delta X_f}{L_i'}\right).
\end{equation}
Here, \(L_i\) is the initial granular column length, \(D_p\) is the mean particle diameter, \(\theta\) is the inclination angle, and \(\Delta X_f\) is the final horizontal runout distance. The effective mobilized length is denoted by \(L_i'\), while \(\nu\) and \(\beta_1\) are the corresponding scaling exponents.

\begin{figure}[h!]
    \centering
    \includegraphics[width=\linewidth]{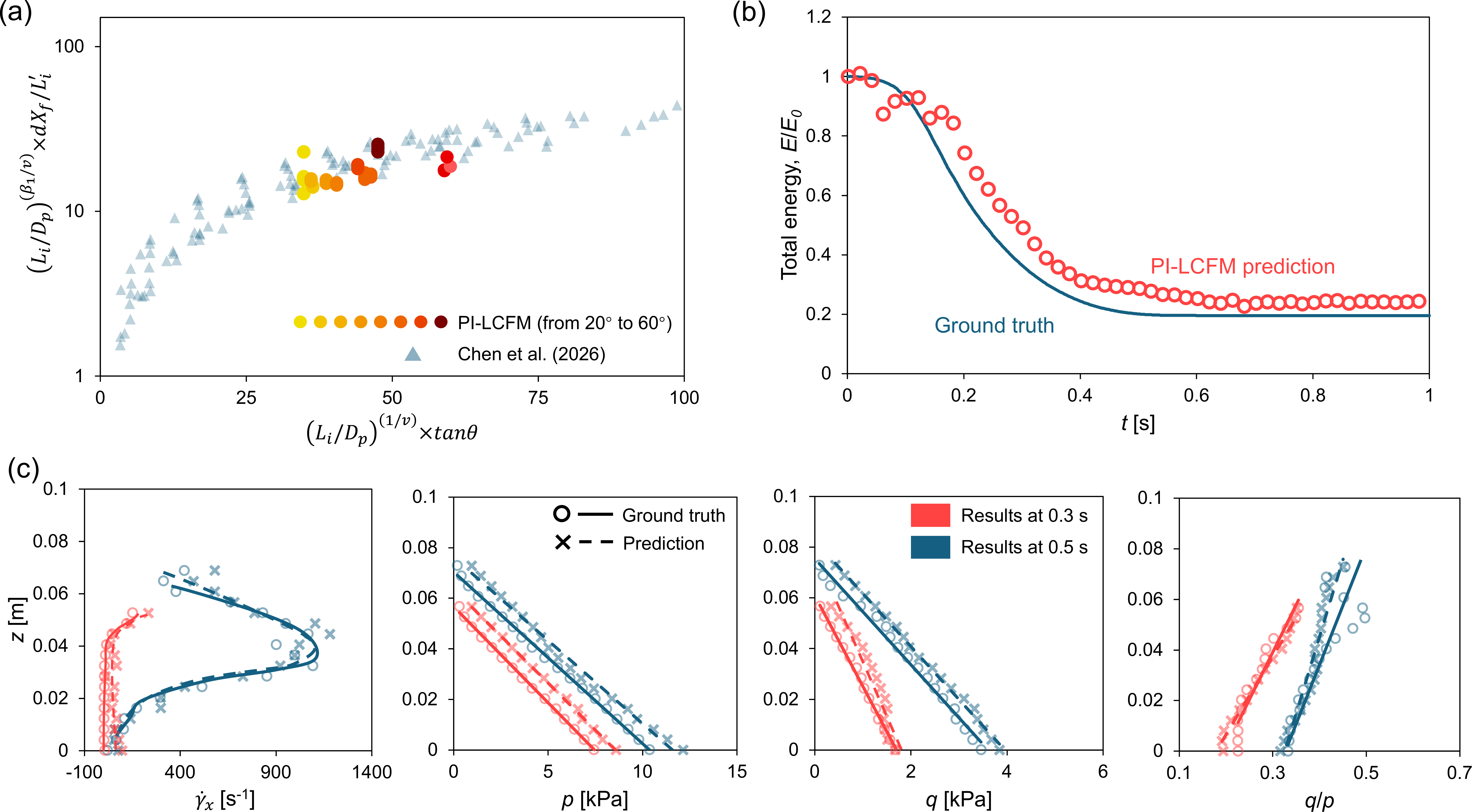}
    \caption{Evaluation of the physical consistency of PI-LCFM predictions. (a) Finite-size scaling of the normalized runout distance, where blue triangles were reported in a previous study \cite{chen2026collapse} and dots are PI-LCFM predictions. (b) Evolution of the total energy normalized by its initial value, \(E(t)/E(0)\). Red circles denote PI-LCFM predictions, while the solid curve represents DEM ground truth. (c) Vertical profiles of the shear rate \(\dot{\gamma}_x\), pressure \(p\), deviatoric stress \(q\), and stress ratio \(q/p\) at \(x=0\) at $t$ = 0.3~s (red) and 0.5~s (blue); circles with solid lines denote the ground truth, and crosses with dashed lines denote PI-LCFM predictions.}
    \label{fig:physics}
\end{figure}

Figure \ref{fig:physics}(a) compares the scaled runout distances predicted from PI-LCFM with the reference data reported in the previous study~\cite{chen2026collapse}. The predicted values follow the reference trend over the range represented by the evaluated cases. This agreement supports the recovery of macroscopic runout scaling behavior, although this scaling relation was not explicitly included in the training objective or inference guidance.
Figure \ref{fig:physics}(b) compares the temporal evolution of the normalized energy, \(E(t)/E(0)\), obtained from PI-LCFM and the DEM reference. The total energy $E(t)$ comprises kinetic, gravitational potential, and granular temperature contributions and was normalized by its initial value $E(0)$. The PI-LCFM prediction captures the rapid energy decrease followed by a slower approach to a nonzero plateau, while the predicted normalized energy remains above the DEM reference during much of the decay and at later times. Nevertheless, the prediction captures the overall timing and magnitude of the energy decay.

Shear-rate and stress profiles provide complementary measures of the local rheological response of granular flows \cite{lu2024shear}. The definitions of $\dot{\gamma}_x$, \(p\), and \(q\) are provided in \ref{app:stress_definitions}. Figure \ref{fig:physics}(c) compares the vertical profiles of the shear rate \(\dot{\gamma}_x\), pressure \(p\), deviatoric stress \(q\), and stress ratio \(q/p\). These quantities were extracted along the \(z\) direction at \(x=0\) at \(t=0.3\) and \(0.5\) s. The predicted profiles reproduce the principal variations in the DEM reference during the two evaluated stages. At \(t=0.3\) s, the MAEs of \(\dot{\gamma}_x\), \(q\), \(p\), and \(q/p\) were \(90.3~\mathrm{s}^{-1}\), \(0.346~\mathrm{kPa}\), \(1.05~\mathrm{kPa}\), and \(0.0263\), respectively. At \(t=0.5\) s, the corresponding MAEs were \(48.3~\mathrm{s}^{-1}\), \(0.254~\mathrm{kPa}\), \(1.06~\mathrm{kPa}\), and \(0.0169\). The lower errors in \(\dot{\gamma}_x\), \(q\), and \(q/p\) at \(t=0.5\) s indicate closer agreement during the later stage, whereas the pressure error remained similar at the two evaluated times. 
Together, these evaluations show that PI-LCFM recovers key granular-flow physical characteristics at both the bulk and local-field levels. Agreement in runout scaling, energy evolution, and stress-related profiles provides evidence of physical fidelity beyond pointwise reconstruction accuracy.

\FloatBarrier

%=============================================================================
\section{Analyses and discussion}

\subsection{Comparison with other AI frameworks}
The performance of PI-LCFM was compared with alternative neural architectures and generative models in terms of reconstruction accuracy and computational cost. Three complementary comparisons were conducted: 1) MLP, temporal MLP, U-Net \cite{ronneberger2015u}, and Temporal DiT (this study) were compared to examine the influence of backbone architecture; 2) PI-LCFM was compared with the score-based denoising diffusion probabilistic model (DDPM) and noise conditional score network (NCSN) models; 3) PI-LCFM was compared with the pixel-space conditional flow matching model developed in our previous study~\cite{li2026generative}. All models were evaluated using the same test data and observation settings within each comparison.
For each model, the best-performing hyperparameter configuration was selected based on validation-set performance under the same model-selection criterion. The selected architectures, training settings, and inference hyperparameters are summarized in \ref{app:baseline_hyperparameters}.

\begin{figure}[h!]
    \centering
    \includegraphics[width=\linewidth]{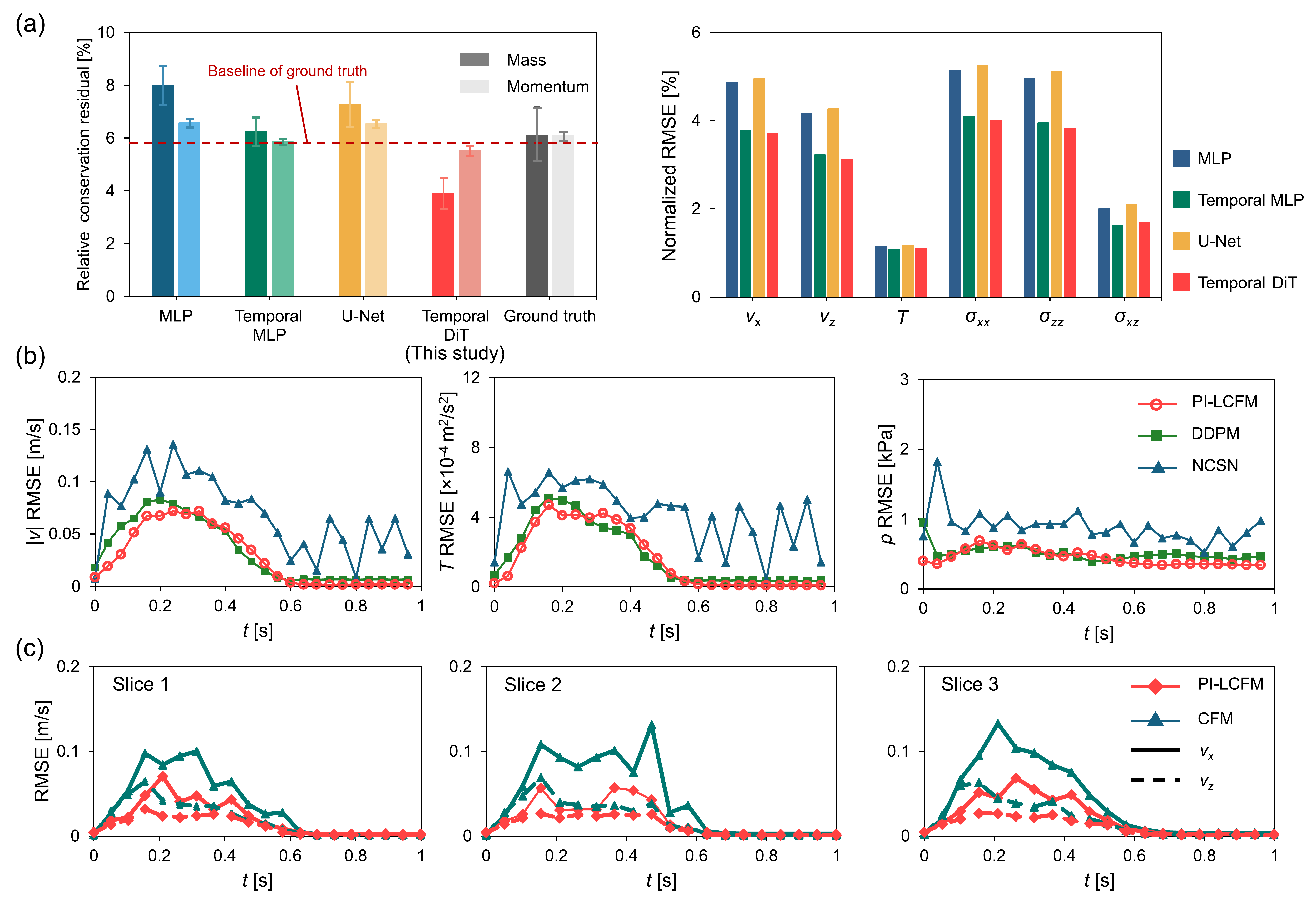}
    \caption{Comparison of PI-LCFM against alternative neural architectures and generative models. 
    (a) Backbone comparison using an MLP, a temporal MLP, a U-Net, and Temporal DiT (this study). The left panel shows relative mass- and momentum-balance residuals (\%), with the DEM results included as a reference. The right panel shows the normalized RMSEs of each reconstructed variable. 
    (b) Time-resolved RMSEs of velocity magnitude, granular temperature, and pressure obtained using PI-LCFM, DDPM, and NCSN. 
    (c) Time-resolved velocity RMSEs obtained using PI-LCFM and the previous frame-by-frame CFM framework \cite{li2026generative} at three interior slices. Solid and dashed curves denote \(v_x\) and \(v_z\), respectively.}
    \label{fig:model_comparison}
\end{figure}

Figure \ref{fig:model_comparison}(a) compares the reconstruction performance of the 4 backbone architectures over different inclination-angle test cases. Physical consistency was evaluated using the mass- and momentum-conservation residuals defined in Eqs. (\ref{eq:mass_residual}-\ref{eq:residuals}), with identical evaluation procedures applied to the reconstructed and DEM fields. 
Reconstruction accuracy was evaluated using the normalized RMSEs of \(v_x\), \(v_z\), \(T\), \(\sigma_{xx}\), \(\sigma_{zz}\), and \(\sigma_{xz}\). The MLP and U-Net produced larger relative conservation residuals than the DEM ground truth, while the temporal MLP achieved residuals close to the reference values. Temporal DiT produced the lowest relative mass and momentum residuals, at 3.9\% and 5.52\%, respectively. It also achieved the lowest or nearly lowest NRMSEs across the 6 reconstructed variables. The stronger performance of the temporal architectures supports the use of temporal dependencies in full-sequence reconstruction.

Figure \ref{fig:model_comparison}(b) compares the time-resolved reconstruction errors of PI-LCFM with two representative score-based generative models, the DDPM and the NCSN \cite{song2020score}. PI-LCFM generally achieved lower RMSEs than both baselines and exhibited substantially smaller temporal fluctuations than NCSN. Its time-averaged RMSEs for velocity magnitude, granular temperature, and pressure were \(0.028~\mathrm{m/s}\), \(1.65\times10^{-4}~\mathrm{m^2/s^2}\), and \(0.44~\mathrm{kPa}\), respectively. The corresponding values were \(0.033~\mathrm{m/s}\), \(1.81\times10^{-4}~\mathrm{m^2/s^2}\), and \(0.51~\mathrm{kPa}\) for DDPM, and \(0.069~\mathrm{m/s}\), \(4.18\times10^{-4}~\mathrm{m^2/s^2}\), and \(0.89~\mathrm{kPa}\) for NCSN. Relative to DDPM, PI-LCFM reduced the time-averaged RMSEs of velocity magnitude, granular temperature, and pressure by \(14.2\%\), \(8.8\%\), and \(12.9\%\), respectively.

The physical consistency of the three models was further assessed by examining global mass and momentum balance. The complete temporal comparisons are provided in Figure \ref{fig:mass_mom_conservation} of \ref{app:mass_mom_conservation}. Among the three generative models, PI-LCFM produced the lowest mass and momentum errors. Compared with DDPM, it reduced the mass and momentum errors by \(5.6\%\) and \(15.3\%\), respectively. Compared with NCSN, the corresponding reductions were \(44.7\%\) and \(34.0\%\). The PI-LCFM mass error remained close to the DEM reference, while its momentum error was lower under the adopted residual metric.

Figure \ref{fig:model_comparison}(c) compares the velocity RMSEs of PI-LCFM and frame-by-frame CFM \cite{li2026generative} at three interior slices. PI-LCFM achieved lower time-averaged errors for both horizontal velocity \(v_x\) and vertical velocity \(v_z\) at every slice, with an overall mean RMSE reduction of  \(45.2\%\). The improvement was greatest during active flow, while both models showed smaller errors after \(t=0.6\) s.
Table~\ref{tab:model_cost_comparison} summarizes the computational costs of generating a 500-frame sequence using an NVIDIA RTX 4090 GPU and an Intel Core i9-13900K processor. With 100 sampling steps, PI-LCFM required \(1.25\) s for unconditional generation and \(43.28\) s for observation-guided reconstruction. 
Including physics guidance increased the PI-LCFM inference time to \(69.19\) s because the physical fields were decoded and the conservation losses were evaluated at each integration step. 
\begin{table}[h!]
    \centering
    \small
    \setlength{\tabcolsep}{6pt}
    \caption{Computational cost of different models}
    \label{tab:model_cost_comparison}
    \begin{tabularx}{\linewidth}{XXXX}
        \toprule
        \multirow{2}{*}{Model} & \multicolumn{3}{c}{Sampling time (s)} \\
        \cmidrule(lr){2-4}
        & Unconditional & Observation-guided & Observation + physics \\
        \midrule
        PI-LCFM  & 1.25 & 43.28 & 69.19 \\
        CFM      & --   & 1905.64& --    \\
        \bottomrule
    \end{tabularx}
\end{table}

The previous frame-by-frame CFM model required \(1905.64\) s for conditional reconstruction of the same number of frames. Each frame required an independent guided ODE integration, whereas PI-LCFM generated the complete latent sequence through one single integration process. Under observation guidance, PI-LCFM achieved an approximately \(44\)-fold speedup relative to CFM. Together with the lower velocity reconstruction errors, this comparison demonstrates the practical benefit of joint latent sequence generation for the evaluated granular flows.
\FloatBarrier

% -----------------------------------------
\subsection{Uncertainty quantification}
Limited observations can admit multiple compatible interior states, motivating an assessment of the variability among reconstructed fields. We characterized the spatial and temporal distribution of ensemble spread and examined the contributions of different sources of variability. Five test cases at plane inclination angles of 20°, 30°, 37.5°, 47.5°, and 55° were evaluated using boundary-velocity observations. For each case, 4 independently trained FM models were combined with 3 realizations of 10\% observation noise and 10 inference initializations. This yielded 120 reconstructed sequences per case and 600 sequences in total. The VAE was fixed throughout this analysis.

\subsubsection{Spatial and temporal distribution of ensemble spread}
\begin{figure}[h!]
    \centering
    \includegraphics[width=\linewidth]{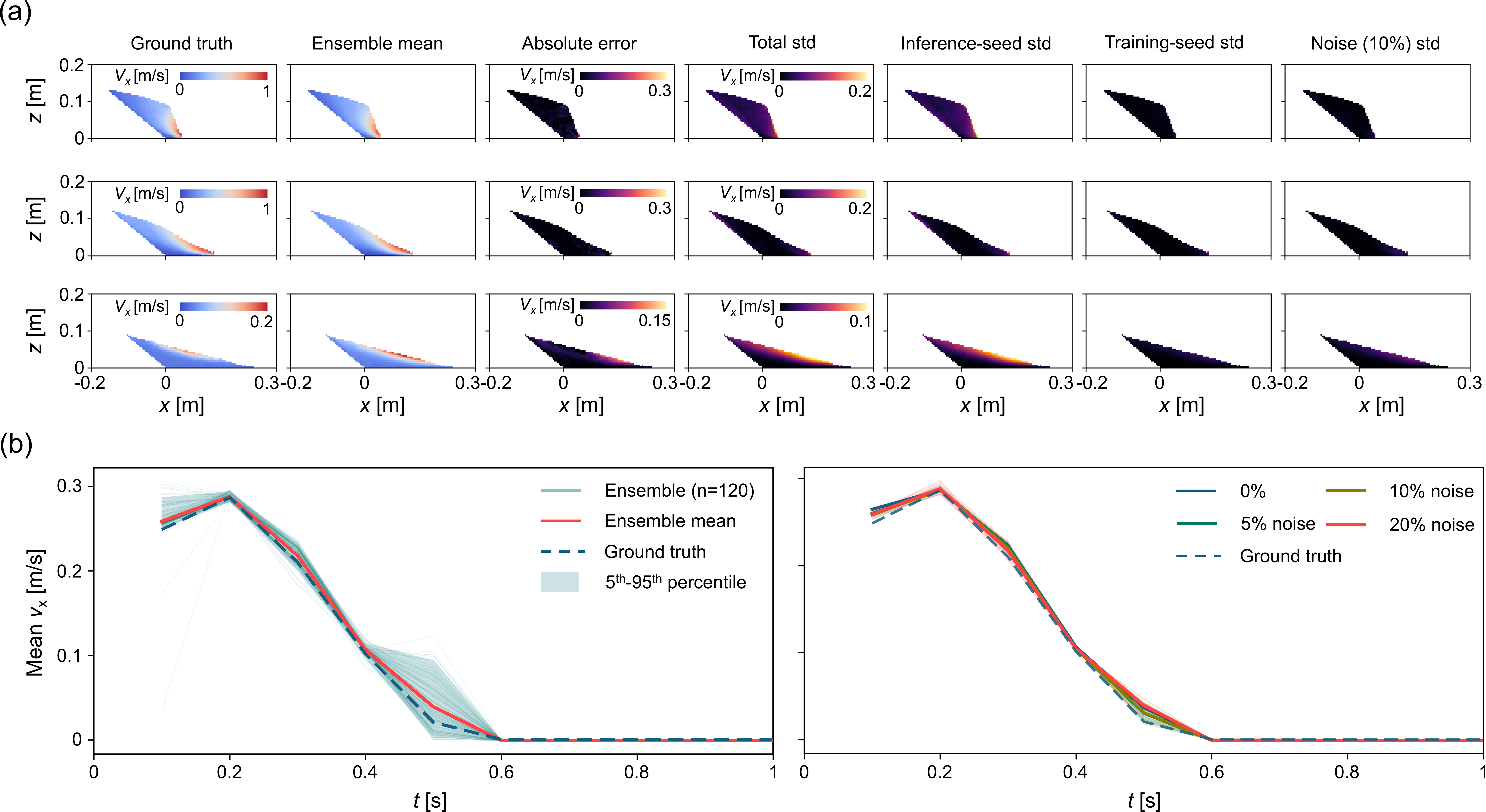}
    \caption{Uncertainty characterization of PI-LCFM for a representative test case.  
    (a) DEM Ground truth, ensemble mean, absolute reconstruction error, total ensemble standard deviation, and component standard deviations associated with inference initialization, FM training, and observation perturbations for \(v_x\) at $t$ = 0.1 s, 0.2 s, and 0.5 s (top to bottom). The component standard deviations are the square roots of the variance components defined in~\ref{app:uq_variance}. 
    (b) Temporal evolution of the domain-averaged \(v_x\) across 120 reconstructions (left) and under different observation-noise levels (right). The shaded region denotes the 5th–95th percentile interval.}
    \label{fig:uq}
\end{figure}

Figure \ref{fig:uq}(a) compares the DEM reference, ensemble mean, absolute reconstruction error, and ensemble standard deviation of \(v_x\) at three representative times. The ensemble mean reproduced the main spatial features of the reference fields. Elevated standard deviations occurred near the advancing flow front and in regions of high velocity or steep velocity gradients. The slowly moving interior exhibited lower ensemble spread. As the material propagated downstream, the region of elevated spread moved with the front and became more localized during deceleration. 

The temporal evolution of the domain-averaged \(v_x\) showed a corresponding pattern (Figure \ref{fig:uq}(b), left). The 5th–95th percentile interval across the 120 reconstructions was widest during active flow and narrowed as the material approached deposition. The ensemble mean followed the DEM trajectory over the same period. These intervals describe the dispersion of the sampled domain-averaged responses.

\subsubsection{Sources of ensemble spread and sensitivity to observation noise}
The ensemble combined variability from inference initialization, observation perturbations, and independent training of the FM model. Table \ref{tab:uq_source_decomposition} summarizes their contributions under the hierarchical variance decomposition defined in \ref{app:uq_variance}. Inference initialization accounted for 80.4\% of the normalized variance on average, followed by model-training variability at 16.4\% and observation perturbations at 3.2\%. The initialization contribution was largest in every case. The training contribution ranged from 4.6\% at 47.5° to 34.9\% at 20°, whereas the observation contribution ranged from 1.9\% to 5.0\%. Thus, within the specified sampling design and decomposition, inference initialization accounted for most of the observed ensemble spread.
\begin{table}[h!]
    \centering
    \small
    \setlength{\tabcolsep}{4.5pt}
    \caption{Hierarchical decomposition of the normalized predictive variance}
    \label{tab:uq_source_decomposition}
    \begin{tabularx}{\linewidth}{XXXXX}
        \toprule
        Case
        & $V_{\mathrm{init}}$ (\%)
        & $V_{\mathrm{obs}}$ (\%)
        & $V_{\mathrm{train}}$ (\%)
        & $V_{\mathrm{total}}$ \\
        \midrule
        $20^{\circ}$   & 62.1 & 3.0 & 34.9 & 0.0088 \\
        $30^{\circ}$   & 85.5 & 1.9 & 12.6 & 0.0170 \\
        $37.5^{\circ}$ & 73.8 & 2.5 & 23.7 & 0.0352 \\
        $47.5^{\circ}$ & 91.6 & 3.7 & 4.6  & 0.0286 \\
        $55^{\circ}$   & 88.9 & 5.0 & 6.1  & 0.0312 \\
        
        \textbf{Case average}
        & \textbf{80.4}
        & \textbf{3.2}
        & \textbf{16.4}
        & \textbf{0.0242} \\
        \bottomrule
    \end{tabularx}
\end{table}

A separate noise-sensitivity analysis examined reconstruction performance as the relative observation-noise level increased from 0\% to 20\% (Figure \ref{fig:uq}(b), right). At the field level, the case-averaged NRMSE of $v_x$ increased by approximately 1.4\% at 20\% noise relative to the noise-free baseline.  These results indicate limited sensitivity of reconstruction accuracy to the tested observation-noise levels.

\FloatBarrier

% -----------------------------------------
\subsection{Ablation study}
Two ablation studies were conducted to examine the effects of conservation guidance and spatial observation coverage on reconstruction performance. The first evaluates whether mass- and momentum-conservation guidance improves the physical consistency of the generated fields without substantially compromising reconstruction accuracy, while the second investigates the amount of spatial observation required for reliable full-field reconstruction.

\subsubsection{Effectiveness of the physics-informed guidance}
The effect of physics-informed guidance was evaluated by comparing conditional reconstructions with and without mass- and momentum-conservation constraints. Each paired comparison used the same pretrained model, observations, numerical settings, and initial Gaussian noise, differing only in the inclusion of physics guidance. The comparison was performed for 6 inclination angles: $20^\circ$, $30^\circ$, $37.5^\circ$, $47.5^\circ$, $55^\circ$, and $60^\circ$. The bulk flow state was assessed using the relative Froude-number error, \(\varepsilon_{\mathit{Fr}}=|\mathit{Fr}-\mathit{Fr}_{\mathrm{DEM}}|/\mathit{Fr}_{\mathrm{DEM}}\), following the procedure in ~\ref{app:froude number}.
Reconstruction accuracy was quantified using the aggregate normalized RMSE averaged equally across the 7 predicted variables, while physical consistency was evaluated using the mass and momentum residuals.

\begin{figure}[h!]
    \centering
    \includegraphics[width=\linewidth]{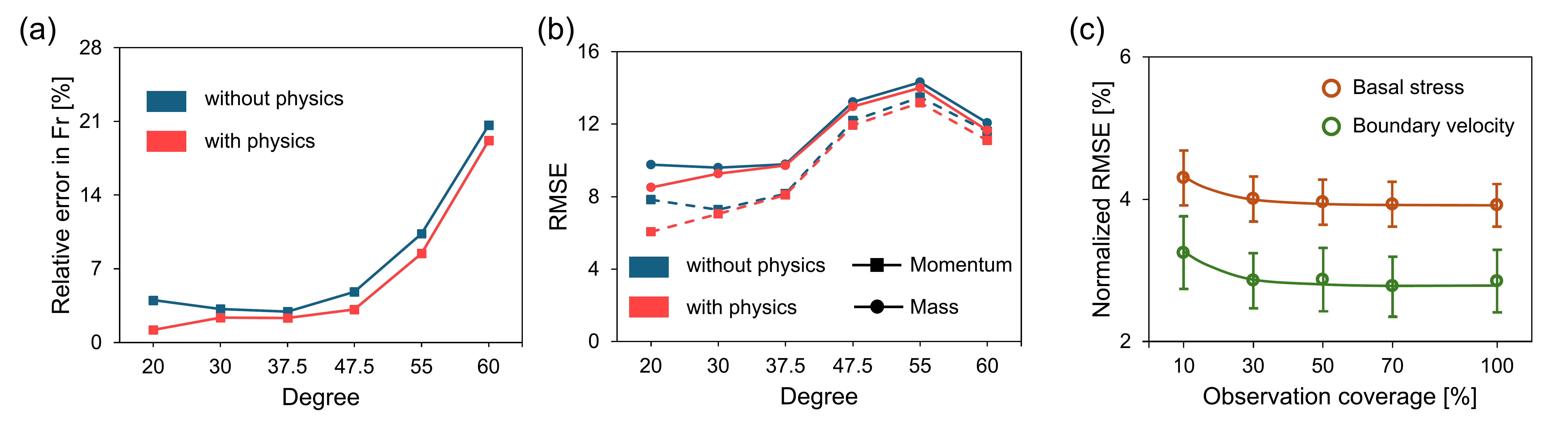}
    \caption{Ablation studies of physics-informed guidance and spatial observation coverage.
    (a) Relative error in the Froude number, $Fr = \frac{\bar{v}}{\sqrt{g h \cos\theta}}$, at different inclination angle with and without physics informed guidance. 
    (b) Corresponding mass- and momentum-conservation residual RMSEs, indicated by circles and squares, respectively. (c) NRMSE averaged across the 7 reconstructed variables at different spatial observation coverages for boundary-velocity and basal-stress observations.}
    \label{fig:ablation}
\end{figure}

As shown in Figure~\ref{fig:ablation}(a), physics guidance reduced the relative Froude-number error at all 6 inclinations, including reductions from 4.0\% to 1.2\% at $20^\circ$ and from 4.8\% to 3.1\% at $47.5^\circ$. These results show that physics guidance improved the reconstructed bulk flow state across the evaluated cases and substantially improved the reconstruction performance.

The conservation residuals exhibited more consistent improvements, as shown in Figure~\ref{fig:ablation}(b). At $20^\circ$, the mass residual decreased from 61.2 to 36.8, and the momentum residual decreased from $9.5\times10^{5}$ to $7.2\times10^{5}$. Improvements were also observed at the larger inclination angles. At $60^\circ$, the mass and momentum residuals decreased by 8.38\% and 6.61\%, respectively. Conservation guidance consistently improves physical consistency while largely preserving field-reconstruction accuracy. 

\subsubsection{Sensitivity to spatial observation coverage} 
The sensitivity of the reconstruction to spatial observation coverage was evaluated separately for boundary-velocity and basal-stress observations, with coverage varied from \(10\%\) to \(100\%\) in each configuration. Reconstruction performance was measured using the full-field, full-sequence normalized RMSE averaged equally across the 7 predicted variables.

Figure~\ref{fig:ablation}(c) shows that increasing the spatial observation coverage generally improved reconstruction accuracy. For boundary-velocity observations, the mean NRMSE decreased from 3.25\% at 10\% coverage to 2.86\%, 2.87\%, and 2.78\% at 30\%, 50\%, and 70\% coverage, respectively. This corresponds to an overall reduction of 0.47\% between 10\% and 70\% coverage. Increasing coverage to \(100\%\) yielded an NRMSE of 2.85\%, indicating no further improvement. The basal-stress configuration exhibited a similar trend and lower mean NRMSEs across the reported coverages. Under the investigated conditions, both configurations showed limited additional benefit beyond approximately \(50\%\)--\(70\%\) coverage. These results suggest that observations over a restricted spatial region can provide sufficient information for full-field reconstruction, whereas extending the measurements over the entire domain yields only a marginal additional improvement under the conditions considered here.

% -----------------------------------------
\subsection{Limitations and future work}
\label{sec:limitations_future}
The framework has been trained and evaluated using DEM-generated data for an inclined-to-horizontal geometry with periodic transverse boundaries. Its performance with experimental measurements remains to be established. Extending the framework to irregular geometry and different basal or lateral boundary conditions will require suitable representations of geometry and boundary conditions, together with evaluation on configurations excluded from training.

The current physics constraints enforce in-plane mass- and momentum-conservation conditions and does not explicitly constrain granular temperature. Future work can extend the guidance to three-dimensional conservation laws and granular-energy constraints involving fluctuation-energy production, transport, and dissipation.

%=============================================================================
\section{Conclusions}
This study developed a physics-informed latent conditional flow matching (PI-LCFM) framework for reconstructing complete spatiotemporal granular-flow fields from partial boundary observations. The framework combines VAE-based dimensionality reduction, temporal flow matching, and training-free observation and physics guidance. This latent formulation enables efficient full-sequence generation while retaining the dominant spatial structures and temporal dynamics of the granular flow. The main conclusions are as follows:

\begin{enumerate}
    \item The latent sequence generation enables efficient reconstruction of granular-flow dynamics. Compared with the frame-by-frame CFM framework, PI-LCFM reduced the velocity-component RMSE averaged across the evaluated interior slices and times by 45.2\%. Observation-guided reconstruction of a 500-frame sequence achieved an approximately 44-fold speedup.

    \item Without retraining, the same pretrained flow-matching prior reconstructs velocity, granular temperature, stress, and solid volume fraction from either partial boundary-velocity or basal-stress observations containing only $\sim 0.3\%$ of the complete field information, with a mean normalized RMSE of 3.18\% over 6 flow variables. Reconstruction of the $60^\circ$, 20,000-particle unseen test case further demonstrated extrapolation beyond the training ranges of both inclination angle and particle number within the prescribed geometry.

    \item Physics guidance reduced mass- and momentum-conservation residuals across the evaluated inclinations while largely preserving or improving field-reconstruction accuracy. The reconstructed sequences also recovered key trends in runout scaling, energy evolution, and vertical shear-rate and stress profiles, providing evidence of physical fidelity beyond pointwise field agreement.

    \item Ensemble spread was concentrated near the advancing flow front and regions of steep velocity gradients, with inference initialization contributing most to the variance under the adopted decomposition. Reconstruction accuracy exhibited limited sensitivity to the tested observation-noise levels.
\end{enumerate}

Overall, PI-LCFM demonstrates that a single reusable latent prior, guided by limited observations and physical constraints, can efficiently reconstruct granular flow physics across measurement modalities without retraining. It also sheds light on exploring internal mechanisms governing flow evolution in applications such as geophysical hazard assessment and industrial process monitoring.

%=============================================================================
\section*{Acknowledgments}
The authors acknowledge the Startup Funds from Texas Tech University (TTU) and the high performance computing resources from the Texas Advanced Computing Center (TACC) at The University of Texas at Austin. The DEM simulations were conducted using MechSys, an open-access multiphysics simulation library, and its source code is available at \url{https://github.com/Axtal/mechsys.git}.

%=============================================================================

%=============================================================================
\section*{CRediT authorship contribution statement}
\noindent
\textbf{Ziqi Han}: Methodology, Validation, Investigation, Software, Visualization, Formal analysis.
\textbf{Kuang Shi}: Data Curation, Software, Writing - Original Draft.
\textbf{Xuyang Li}: Writing - Review \& Editing.
\textbf{Teng Man}: Software, Writing - Review \& Editing.
\textbf{Yimin Lu}: Conceptualization, Supervision, Resources, Funding acquisition, Visualization, Writing - Review \& Editing.

%=============================================================================
\bibliography{references}

\newpage

\appendix
\counterwithin{table}{section}
\renewcommand{\thetable}{\Alph{section}-\arabic{table}}
\section{Coarse-graining and stress-field smoothing}
\label{app:coarse_graining}
The particle-scale quantities were mapped onto a structured Eulerian grid. For each cell, the velocity was calculated by averaging the velocities of particles whose centers were located within the cell. The granular temperature was evaluated from the particle velocity fluctuations relative to the cell-averaged velocity.

The raw cell-scale contact stress $\overline{\sigma}_{ij}$ was calculated using Eq.~\eqref{eq:cell_contact_stress}. Because the number and configuration of particle contacts vary between adjacent cells, the raw stress fields contain local particle-scale fluctuations. The stress tensor was therefore spatially averaged over a neighborhood $\mathcal{N}_{m,n}$ surrounding cell $(m,n)$:
\begin{equation}
\widetilde{\sigma}_{ij}^{(m,n)}
=
\frac{1}{N_{\mathcal{N}}}
\sum_{(a,b)\in\mathcal{N}_{m,n}}
\overline{\sigma}_{ij}^{(a,b)},
\label{eq:stress_spatial_average}
\end{equation}
where $\widetilde{\sigma}_{ij}^{(m,n)}$ is the smoothed stress component at cell $(m,n)$, $\mathcal{N}_{m,n}$ denotes the set of neighboring cells included in the averaging operation, and $N_{\mathcal{N}}$ is the number of cells in that neighborhood. This operation reduces localized fluctuations caused by the discrete particle contacts while retaining the dominant spatial distribution of the stress field.

\section{Dataset conditions}
\label{app:dataset_conditions}
The dataset contained 88 DEM simulation cases generated using different inclination angles, particle numbers, and random initialization seeds. Each case contained 500 snapshots recorded at intervals of 0.002 s, corresponding to a total simulated duration of 1.0 s. The simulation conditions are summarized in Table~\ref{tab:dataset_conditions}. 
\begin{table}[h!]
    \centering
    \caption{Summary of the DEM simulation conditions used to construct the dataset}
    \label{tab:dataset_conditions}
    \begin{tabular}{cccc}\toprule
        
        Inclination angle ($^\circ$)
        & Particle number $N_p$
        & Random seeds
        & Number of cases \\\midrule
        
        15, 20, 25
        & 40,000
        & 500, 1000, 2000& 9 \\

        30, 35, 37.5, 40, 42.5, 45, 47.5, 50
        & 40,000, 50,000, 60,000
        & 500, 1000, 2000& 72 \\

        55
        & 40,000, 50,000
        & 500, 1000, 2000& 6 \\

        60
        & 20,000& 500& 1 \\
        
        \multicolumn{3}{c}{Total}
        & 88 \\ \bottomrule
        
    \end{tabular}
\end{table}

\begin{table}[h!]
    \centering
    \caption{Test cases in dataset}
    \label{tab:test_cases}
    \begin{tabularx}{\linewidth}{XXXX}
        \toprule
        Number& Angle ($^\circ$) & $N_p$ & Seeds
        \\
        \midrule
        1, 2, 3& 20   & 40,000 & 500, 1000, 2000
        \\
        
        4& 25   & 40,000 & 500
        \\
        
        5& 30   & 40,000 & 2000
        \\
        
        6& 35   & 40,000 & 500
        \\
        
        7& 37.5 & 40,000 & 1000
        \\
        
        8, 9, 10& 40   & 50,000 & 500, 1000, 2000
        \\
        
        11& 45   & 40,000 & 1000
        \\
        
        12& 47.5 & 60,000 & 500
        \\
        
        13& 50   & 60,000 & 500
        \\
        
        14& 55   & 50,000 & 500
        \\
        
        15& 60   & 20,000 & 500
        \\ \bottomrule
    \end{tabularx}
\end{table}
\FloatBarrier

\clearpage

\section{Hyperparameter of network}
\label{app:hyperparameter_network}
\begin{table} [h!]
    \centering
    \caption{Summary of network architecture and training hyperparameter.}
    \label{tab:hyperparams}
    \begin{tabularx}{\linewidth}{XXX}
        \toprule
        Module & Hyperparameter & Value \\
        \midrule
        \multirow{4}{*}{VAE} & Latent dimension $d$ & 256\\
         & KL weight $\beta$ & 1e-04\\
         & Batch size & 512 \\
         & Initial learning rate & 1e-03\\
        
        \multirow{6}{*}{Temporal DiT (FM)}& Hidden dimension & 768 \\
         & Transformer blocks & 8 \\
         & Attention heads & 8 \\
         & MLP expansion ratio & 4 \\
         & Batch size & 128\\
         & Initial learning rate & $1\times10^{-3}$ \\
        
        \multirow{2}{*}{Guidance sampling}& Integration steps & 100 \\
         & \(\alpha_m\), \(\alpha_p\)& 1, 1\\ \bottomrule
    \end{tabularx}
\end{table}

\FloatBarrier

\section{Definitions of pressure, deviatoric stress, and shear rate}
\label{app:stress_definitions}
The pressure $p$ and deviatoric stress $q$ were calculated from the stress components as,
\begin{equation}
p=\frac{\widetilde{\sigma}_{xx}+\widetilde{\sigma}_{zz}}{2},
\qquad
q=\sqrt{
\left(\frac{\widetilde{\sigma}_{xx}-\widetilde{\sigma}_{zz}}{2}\right)^2
+\widetilde{\sigma}_{xz}^{\,2}
}
\end{equation}
Here, $p$ represents the mean normal stess, and $q$ is the maximum deviatoric stess in the $x-z$ plane. Compressive normal stresses are taken as positive.

The shear rate \(\dot{\gamma}_x\) was defined as the vertical gradient of the horizontal velocity,
\begin{equation}
    \dot{\gamma}_x=\frac{\partial v_x}{\partial z},
\end{equation}
where $v_x$ is the horizontal velocity and $z$ is the vertical coordinate. The shear rate has units of $s^{-1}$.

\section{Reconstruction from partial basal stress observations}
\label{app:basal_stress}

\begin{figure}[htbp]
    \centering
    \includegraphics[width=\linewidth]{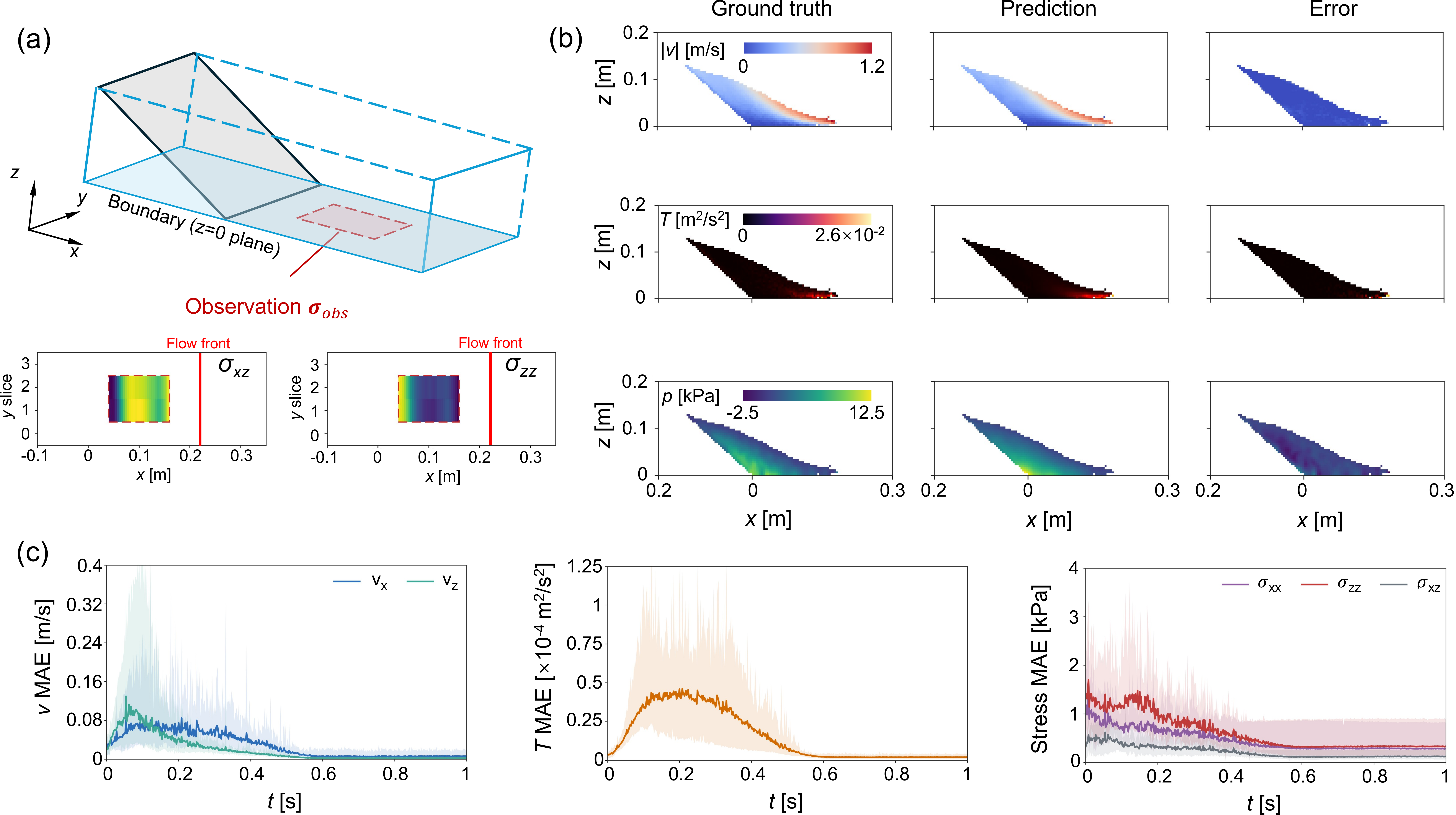}
    \caption{Conditional generation from partial basal stress observations. (a) Schematic view of partial stress observation: two basal stress component $\sigma_{xz}$ and $\sigma_{zz}$ are observed within a rectangular window (red dashed box) on the bottom. 
(b) Ground truth, GenAI prediction, and absolute error contour of the velocity magnitude $|v|$, granular temperature $T$ , and pressure $p$. 
(c) Temporal evolution of the mean absolute error (MAE) of individual velocity components, granular temperature, and stress components; solid lines denote mean over all test cases and shaded bands indicate the range across all test cases. } \label{fig:partial_obs_sigma}
\end{figure}

Figure~\ref{fig:partial_obs_sigma} provides supplementary results for reconstruction from partial basal stress observations. Panel (a) illustrates the observation configuration. Panel (b) compares the DEM reference, prediction, and error fields for the representative $40^\circ$ case at $t=0.24$ s. Panel (c) summarizes the temporal reconstruction errors across 15 cases under the same observation configuration.
\FloatBarrier

\newpage
\section{Hyperparameters of the comparison models}
\label{app:baseline_hyperparameters}
\begin{table*}[htbp]
\centering
\small
\caption{Hyperparameter configurations adopted for the comparison models}
\label{tab:baseline-hyperparameters}

\begin{adjustbox}{width=\textwidth}
\begin{tabular}{lcccc}
\toprule
Model
& Main architecture
& Batch size
& Learning rate
& Sampling steps
\\
\midrule

MLP
& Hidden dim.\ 512, 4 blocks
& 128
& $1\times10^{-3}$
& 100
\\

Temporal MLP
& Hidden dim.\ 768, 4 blocks
& 128
& $1\times10^{-3}$
& 100
\\

U-Net
& Channels 64--128--256--512
& 32
& $1\times10^{-3}$
& 100
\\

Temporal DiT
& Hidden dim.\ 768, 8 blocks, 8 heads
& 128
& $1\times10^{-3}$
& 100
\\

DDPM
& \shortstack{U-Net, base channels 32,\\
  multipliers 1--2--2--4, 2 residual blocks}
& 128& $2\times10^{-4}$
& 100
\\

NCSN
& \shortstack{Base channels 32, multipliers 1--2--2--4,\\
  2 residual blocks, InstanceNorm++}
& 128& $2\times10^{-4}$
& 100
\\

CFM
& \shortstack{U-Net, channels 128--256--512--512,\\
  2 residual blocks per level}
& 16
& {$2.5\times10^{-5}$}& \shortstack{100 time points per frame\\
  (99 Euler updates)}
\\

\bottomrule
\end{tabular}
\end{adjustbox}
\end{table*}

\FloatBarrier

\section{Conservation comparison with score-based generative models}
\label{app:mass_mom_conservation}
\begin{figure}[htbp]
    \centering
    \includegraphics[width=\linewidth]{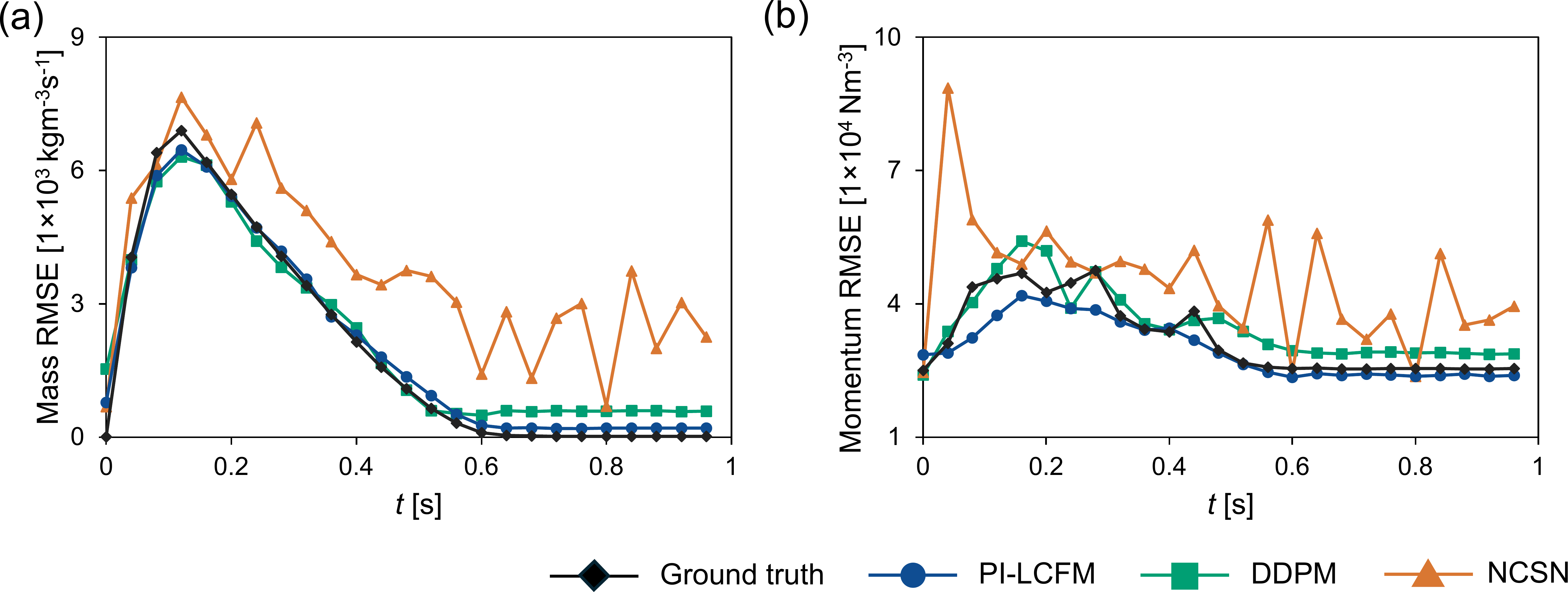}
    \caption{Temporal evolution of conservation-residual RMSEs obtained from PI-LCFM, DDPM, NCSN, and the DEM reference. (a) Mass conservation and (b) Momentum conservation.} \label{fig:mass_mom_conservation}
\end{figure}

The temporal evolution of the globally integrated mass and momentum was compared among PI-LCFM, DDPM, NCSN, and the DEM reference. As shown in Figure \ref{fig:mass_mom_conservation}, PI-LCFM followed the DEM reference more closely than the two score-based baselines over most of the sequence. NCSN exhibited the largest deviations and temporal fluctuations, whereas PI-LCFM produced the lowest aggregate conservation errors among the evaluated generative models.

\section{Hierarchical decomposition of ensemble variance}
\label{app:uq_variance}

For each test case, the ensemble contained $M=4$ independently trained FM models, $O=3$ observation-noise realizations, and $S=10$ inference initializations. Let $x_{mos}^{(v)}(u)$ denote the normalized reconstructed value of variable \(v\) at a spatial–temporal location \(u\). The indices $m$, $o$, and $s$ identify the model, observation realization, and inference initialization, respectively. The case index is omitted for clarity. The VAE was fixed throughout the ensemble.

The decomposition follows the law of total variance, conditioning first on the model and then on the observation realization:
\begin{equation}
\begin{aligned}
\operatorname{Var}(x)
={}&\mathbb{E}_{m,o}\!\left[\operatorname{Var}_{s}(x\mid m,o)\right] \\
 &+\mathbb{E}_{m}\!\left[\operatorname{Var}_{o}\!\left(\mathbb{E}_{s}[x\mid m,o]\mid m\right)\right] \\
 &+\operatorname{Var}_{m}\!\left(\mathbb{E}_{o,s}[x\mid m]\right).
\end{aligned}
\label{eq:uq_total_variance_law}
\end{equation}
Here, expectations and variances are taken over the equally weighted empirical ensemble. The three terms quantify variation across inference initializations, observation realizations within each model, and model-specific ensemble means.

At each location and for each variable, define the successive ensemble means as
\begin{equation}
\bar{x}_{mo}=\frac{1}{S}\sum_{s=1}^{S}x_{mos},\qquad
\bar{x}_{m}=\frac{1}{O}\sum_{o=1}^{O}\bar{x}_{mo},\qquad
\bar{x}=\frac{1}{M}\sum_{m=1}^{M}\bar{x}_{m},
\label{eq:uq_ensemble_means}
\end{equation}
where $v$ and $u$ have been suppressed. The empirical variance components are
\begin{align}
v_{\mathrm{init}}
&=\frac{1}{MOS}\sum_{m=1}^{M}\sum_{o=1}^{O}\sum_{s=1}^{S}
\left(x_{mos}-\bar{x}_{mo}\right)^2,
\label{eq:uq_init_variance}\\
v_{\mathrm{obs}}
&=\frac{1}{MO}\sum_{m=1}^{M}\sum_{o=1}^{O}
\left(\bar{x}_{mo}-\bar{x}_{m}\right)^2,
\label{eq:uq_obs_variance}\\
v_{\mathrm{train}}
&=\frac{1}{M}\sum_{m=1}^{M}
\left(\bar{x}_{m}-\bar{x}\right)^2.
\label{eq:uq_train_variance}
\end{align}
These components satisfy the exact empirical identity
\begin{equation}
v_{\mathrm{total}}
=\frac{1}{MOS}\sum_{m=1}^{M}\sum_{o=1}^{O}\sum_{s=1}^{S}
\left(x_{mos}-\bar{x}\right)^2
=v_{\mathrm{init}}+v_{\mathrm{obs}}+v_{\mathrm{train}}.
\label{eq:uq_variance_sum}
\end{equation}
The identity follows by expanding $x_{mos}-\bar{x}$ into three successive deviations,
\begin{equation}
x_{mos}-\bar{x}
=\left(x_{mos}-\bar{x}_{mo}\right)
+\left(\bar{x}_{mo}-\bar{x}_{m}\right)
+\left(\bar{x}_{m}-\bar{x}\right).
\end{equation}
All cross terms vanish after summation. The variance denominators use the numbers of ensemble members at each level, without Bessel corrections. This convention preserves the additive identity for the finite ensemble. The attribution depends on the conditioning order, with interaction effects included in the corresponding conditional components.

The case-level variance was obtained by averaging the variance fields over the evaluated spatial and temporal locations and variables:
\begin{equation}
V_k=\frac{1}{C}\sum_{v=1}^{C}
\frac{1}{|\Omega_v|}\sum_{u\in\Omega_v}
{v_k^{(v)}(u)},
\qquad k\in\{\mathrm{init},\mathrm{obs},\mathrm{train},\mathrm{total}\}.
\label{eq:uq_normalized_variance}
\end{equation}
Here, $C$ is the number of evaluated variables, and $\Omega_v$ is the set of evaluated spatial--temporal locations for variable $v$. The same evaluation locations were used for all ensemble members
and variance components within each case.
The same evaluation locations and normalization scales are used for all ensemble members and all variance components within a case. Consequently, $V_{\mathrm{total}}=V_{\mathrm{init}}+V_{\mathrm{obs}}+V_{\mathrm{train}}$. The percentage contribution of each component is calculated after aggregation:
\begin{equation}
p_k=100\frac{V_k}{V_{\mathrm{total}}},
\qquad k\in\{\mathrm{init},\mathrm{obs},\mathrm{train}\}.
\label{eq:uq_variance_percentage}
\end{equation}
For the five cases, the final table row reports the arithmetic mean of the case-specific percentages and the arithmetic mean of the total normalized variances. The decomposition quantifies ensemble variability under the specified hierarchy. Reconstruction errors associated with the fixed VAE are evaluated separately against the DEM reference.

\section{Froude number computation}
\label{app:froude number}
The bulk mobility of the flow was characterized by the Froude number,
\begin{equation}
  \mathit{Fr} = \frac{\bar{u}}{\sqrt{g\,h\cos\theta_b}}
  \label{eq:froude}
\end{equation}
where $\bar{u}$ is the depth-averaged flow speed of an $x$-column weighted by the solid volume fraction, $h$ is the bed-normal flow depth, $g$ is the gravitational acceleration, and $\theta_b$ is the local bed inclination ($\theta$ on the inclined plane and $0$ on the horizontal plane). Only cells with $\phi > 0.0327$ were included, and columns containing at least two such cells were retained. The column values were averaged with the column mass as the weight over the four slices and over the active stage of the flow, defined as the frames in which the total kinetic energy of the DEM reference exceeds 5\% of its peak value. The same procedure was applied to the reconstructed and DEM fields, and the relative error $\varepsilon_{\mathit{Fr}} = |\mathit{Fr}-\mathit{Fr}_{\mathrm{DEM}}|/\mathit{Fr}_{\mathrm{DEM}}$ was used to assess whether a reconstruction recovers the bulk flow state.
\FloatBarrier
\end{document}